# Incomplete carbon-oxygen detonation in Type Ia supernovae

Inma Domínguez

*Departamento de Física Teórica y del Cosmos,*
*University of Granada, 18071 Granada, Spain*

and

Alexei Khokhlov

*Department of Astronomy & Astrophysics*
*and the Enrico Fermi Institute,*
*The University of Chicago, Chicago, IL, 60637*

## ABSTRACT

Incomplete carbon-oxygen detonation with reactions terminating after burning of $C^{12}$ in the leading $C^{12}+C^{12}$ reaction ($C$-detonation) may occur in the low density outer layers of white dwarfs exploding as Type Ia supernovae (SN Ia). Previous studies of carbon-oxygen detonation structure and stability at low densities were performed under the assumption that the *velocity* of a detonation wave derives from complete burning of carbon and oxygen to iron. In fact, at densities $\rho \leq 10^6$ g/cm$^3$ the detonation in SN Ia may release less than a half of the available nuclear energy. In this paper we study basic properties of such detonations. We find that the length of an unsupported steady-state $C$-detonation is $\simeq 30-100$ times greater than previously estimated, and that the decreased energy has a drastic effect on the detonation stability. In contrast to complete detonations which are one-dimensionally stable, $C$-detonations may be one-dimensionally unstable and propagate by periodically re-igniting themselves via spontaneous burning. The re-ignition period at $\rho \leq 10^6$ g/cm$^3$ is estimated to be greater than the time-scale of a SN Ia explosion. This suggests that propagation and quenching of $C$-detonations at these densities could be affected by the instability. Potential observational implications of this effect are discussed.

*Subject headings:* hydrodynamics - instabilities - nuclear reactions, nucleosynthesis, abundances - shock waves - supernovae: general - white dwarfs

## 1. Introduction

Type Ia Supernovae (SNIa) are thermonuclear explosions of carbon-oxygen white dwarfs (CO-WD) in binary stellar systems. Explosion models of SNIa currently discussed in the literature include explosions of Chandrasekhar-mass WD, such as delayed detonation, pulsating delayed detonations and its variants (Khokhlov 1991a,b; Gamezo et al. 2005; Livne et al. 2005; Röpke & Niemeyer 2007; Jackson et al. 2010), gravitationally confined (Plewa et al. 2004; Plewa 2007; Jordan et al. 2008; Meakin et al. 2009) and pulsating reverse detonations (Bravo & García-Senz 2006; Bravo et al. 2009; Bravo & García-Senz 2009), explosions of sub-Chandrasekhar WD (Woosley et al. 1980; Nomoto 1982; Livne & Glasner 1991; Livne & Arnett 1995; Fink et al. 2007, 2010) and super-massive WD (Pfannes et al. 2010). These models differ in their assumptions about initial conditions, ignition processes, whether the explosion involves subsonic deflagration or not, and other details, and they have a varying success in explaining basic observations of SN Ia (this subject is outside of the scope of the paper). A common feature of the models is that all of them involve, in one way or another, the detonation mode of burning.

Detonation is a supersonic wave of burning in which reactions are triggered by a strong shock. The energy released by burning maintains the shock strength and propagation of the detonation wave. A steady-state detonation is described by the ZND theory (Zeldovich 1940; von Neumann 1942; Döring 1943) according to which the detonation wave consists of a leading planar shock followed by a one-dimensional reaction zone. A ZND detonation structure is usually unstable due to a strong positive feedback between hydrodynamical perturbations and energy release inside the reaction zone, and a real detonation propagates non-steadily with strong oscillations and transverse waves which form a multi-dimensional cellular structure of a detonation wave (e.g. Fickett & Davis 1979; Lee 1984; Shepherd 2009).

Explosive burning in SN Ia consists of three distinct stages - burning of carbon to $O$, $Ne$, $Na$, $Mg$ and some $Si$ ($C$-burning), subsequent burning of oxygen and formation of $Si$-group elements ($O$-burning), and finally burning of silicon to $Fe$-group elements ($Si$-burning). Nuclear statistical equilibrium (NSE) in burned matter sets in at the end of $Si$-burning. Time-scales of $C$, $O$, and $Si$-burning, $t_C$, $t_O$ and $t_{Si}$, differ by orders of magnitude, $t_C \ll t_O \ll t_{Si}$, and increase exponentially with decreasing temperature. Due to the existence of three distinct stages of burning, a CO detonation waves in SN Ia consists of $C$, $O$, and $Si$-burning layers of increasing thickness, $x_C \ll x_O \ll x_{Si}$, following each other. All three thicknesses depend on density, $\rho$, and increase exponentially with decreasing $\rho$. Thickness of $Si$ and $O$-burning layers, $x_{Si}$ and $x_O$, become greater than the characteristic scale of an exploding WD ($\simeq 10^8$ cm) at $\rho \leq 10^7$ g/cm$^3$ and $\rho \leq 10^6$ g/cm$^3$, respectively (e.g. Khokhlov 1989; Gamezo et al. 1999; Sharpe 1999). Equivalently, it could be said that at these densities $t_{Si}$ and $t_O$ become greater than the characteristic explosion time-scale of WD $\simeq 1$ sec. An important observational property of SN Ia, as was early inferred from near-maximum spectra, is the presence of intermediate mass elements such as $Si$, $S$, $Ca$ $Mg$, $Na$ and $O$ in outer layers of a SN Ia envelope (e.g. Pskovskii 1969; Branch et al. 1982, 1983). Presence of these elements can be explained as a consequence of incomplete burning at low densities $\rho \leq 10^7$ g/cm$^3$.

Whether and how the instability and cellular structure of a detonation influences the explosion of a SN Ia has been a long-standing question. Time-dependent interplay of nuclear reactions and hydrodynamical motions in a detonation wave may affect explosive nucleosynthesis and quenching of burning. Incomplete $C$-burning may leave some carbon intact. This may be important for prediction of chemical composition and formation of light curves and spectra of SN Ia before and near maximum light. Due to the resolution limitations these effects were studied in idealized constant-density conditions representing small parts of an exploding white dwarf. A one-dimensional longitudinal instability of $CO$ detonations was studied numerically in Khokhlov (1993) (see also Koldoba et al. (1994)). Cellular structure of CO detonations was obtained in numerical simulations of Boisseau et al. (1996); Gamezo et al. (1999); Timmes et al. (2000). Calculations show that cellular structure associated with $O$ and $Si$-burning might be potentially important and affect burning at densities $\rho \leq 10^7$ g/cm$^3$ but cellular structure related to $C$-burning appears to be very small at all densities of interest, say, at $\rho > 10^5$ g/cm$^3$.

For low-density detonations the simulations mentioned above contain a serious deficiency. The reason is as follows. The simulations use background conditions (density, temperature, etc.) and the detonation velocity as input parameters. These parameters are used to establish initial and boundary conditions for a simulation. The simulation then gives information about scales and stability of a detonation wave. In the above simulations it was assumed that the input detonation velocity corresponds to a complete burning to NSE. For detonations at low density this leads to a contradiction: $Si$, and $O$-burning at low densities may not occur and a detonation wave may consist of a shock wave followed either by $C$ and $O$-burning layers ($O$-detonation) or by a $C$-burning layer alone ($C$-detonation). Nuclear energy released in such detonation waves will be smaller than the NSE energy release. It is this reduced energy release which must be used for calculating the input detonation velocity. The assumption of NSE will lead to a higher detonation velocity, over-prediction of post-shock temperature and density, under-estimation of a detonation wave thickness, and may affect detonation stability and propagation.

In this paper we re-visit the problem of detonation structure and stability taking into account the reduction of energy release in incomplete detonation waves. We try to answer the following questions: What are the basic properties and characteristic spatial and temporal scales of incomplete $O$ and $C$-detonation waves at low densities? How unstable are these detonations? What is the time-scale of the instability and how does it compare to other scales of the problem? To what extent may the instability affect quenching of a detonation wave and the resulting nucleosynthesis? In this paper we study ZND structure and one-dimensional stability of incomplete detonations. Propagation of a detonation wave is also influenced by multi-dimensional effects but a one-dimensional investigation should be the first step. We find in this paper that a $C$-detonation could be a factor of $\simeq 30-100$ thicker than previously thought and that $C$-detonations could be highly unstable with respect to one-dimensional longitudinal perturbations. Multi-dimensional study of $C$-detonations will be presented in the next paper.

The paper is organized as follows. Sect. 2 present governing equations and input physics. Sect. 3 discusses properties of steady-state detonations with complete and partial energy release. Sect. 4 describes numerical simulations and presents the analysis of one-dimensional stability of the detonations. The results and potential implications for SN Ia modeling and observations are discussed in Sect. 5. Numerical method and test simulations are described in Appendix A.

## 2. Formulation of the problem

Propagation of a one-dimensional detonation wave is described by the time-dependent, compressible, reactive flow Euler equations of fluid dynamics,

$$\frac{\partial \rho}{\partial t} = -\frac{\partial (\rho u)}{\partial x}, \tag{1}$$

$$\frac{\partial \rho u}{\partial t} = -\frac{\partial (\rho u^2 + P)}{\partial x}, \tag{2}$$

$$\frac{\partial E}{\partial t} = -\frac{\partial (u(E+P))}{\partial x} + \rho \dot{q}, \tag{3}$$

$$\frac{\partial \rho \bar{Y}}{\partial t} = -\frac{\partial (\rho u \bar{Y})}{\partial x} + \rho \bar{R}, \tag{4}$$

for mass density $\rho$, velocity $u$, energy density $E = \rho(\epsilon + \frac{u^2}{2})$, where $\epsilon$ is internal energy per unit mass, and the composition of reactants $\bar{Y} = \{Y_1, ..., Y_N\}$ defined as $Y_i = n_i/\rho N_a$, where $N$ is the number of reactants, $n_i$ - their number densities, and $N_a$ is the Avogadro number. $\bar{R} = \{R_1, ..., R_N\}$ are the corresponding net reaction rates which are functions of $\bar{Y}$, $\rho$, and temperature $T$. The energy generation rate due to nuclear reactions is

$$\dot{q} = \bar{Q} \cdot \bar{R} = \sum_{i=1}^{N} Q_i R_i, \quad (5)$$

where $\bar{Q} = \{Q_1, ..., Q_N\}$ are the binding energies of nuclei. Numerical method of integration of equations (1) - (4) is described in Appendix A.

Nuclear kinetics is described by the $\alpha$-network which has been extensively used in previous studies of detonation stability and cellular structure (Khokhlov 1993; Gamezo et al. 1999; Timmes et al. 2000). The network consists of $N = 13$ nuclei $He^4$, $C^{12}$, $O^{16}$, $Ne^{20}$, $Mg^{24}$, $Si^{28}$, $S^{32}$, $Ar^{36}$, $Ca^{40}$, $Ti^{44}$, $Cr^{48}$, $Fe^{52}$, and $Ni^{56}$. The network takes into account binary reactions between $\alpha$-particles and heavier nuclei $C^{12}+\alpha \leftrightarrow O^{16}+\gamma$, $O^{16}+\alpha \leftrightarrow Ne^{20}$, ..., $Fe^{52} + \alpha \leftrightarrow Ni^{56}$; $C^{12} + C^{12}$, $C^{12} + O^{16}$, and $O^{16} + O^{16}$ and a triple-alpha reaction $3\alpha \leftrightarrow C^{12} + \gamma$. Effective reaction rates of binary reactions involving $\alpha$ particles are calculated as the sums of contributions of $(\alpha,p)(p,\gamma)$, $(\alpha,n)(n,\gamma)$, $(\alpha,\gamma)$ reaction channels involving $p$, $n$, and $\gamma$-photons. Forward reaction rates and partition functions are taken from compilations of Fowler et al. (1978); Woosley et al. (1978); Thielemann (1993). Reverse reaction rates are calculated from the principle of detailed balance. The network captures the multi-stage nature of explosive CO burning and correctly reproduces the time-scales of C, O, Si burning stages, and the onset of NSQE and NSE in nuclear matter. The equation of state includes contributions from ideal Fermi-Dirac electrons and positrons with arbitrary degeneracy and relativism, equilibrium Planck radiation, and ideal Boltzmann nuclei.

## 3. Steady-state CO detonations

### 3.1. ZND detonation in CO mixtures

Steady-state detonation solutions provide basic temporal and spatial scales of a detonation wave are also used as initial conditions for time-dependent numerical simulations. In a reference frame moving with the detonation velocity, $D$, all variables inside a one-dimensional steady-state detonation wave are a function of the distance, $x$, from the leading shock and do not depend on $t$. We denote with subscripts 0, $s$, and $d$ hydrodynamical variables in unburned cold matter, in unburned matter located immediately behind the leading shock, and in detonation products, respectively. Hydrodynamical variables inside the reaction zone are related to hydrodynamical variables in unburned matter through the Hugoniot relations,

$$\rho u = \rho_0 D, \quad (6)$$

$$P - P_0 = -(\rho_0 D)^2 \cdot (V - V_0), \quad (7)$$

$$\epsilon - \epsilon_0 + \frac{1}{2}(P + P_0) \cdot (V - V_0) + q = 0, \tag{8}$$

where $V = 1/\rho$ is the specific volume, and

$$q = \bar{Q} \cdot (\bar{Y}_0 - \bar{Y}) \tag{9}$$

is the energy released in a fluid element by burning from 0 to $x$. On a $P$-$V$ diagram (Fig. 1) physical conditions of matter inside the reaction zone are located on the Rayleigh-Mikhelson (R-M) line (7) with the slop $tg\,\alpha = -(\rho_0 D)^2$. For $q = 0$, (8) gives the usual shock adiabat, $S$. Post-shock conditions are determined by the intersection of the R-M line with $S$. For $q = q_d$, (8) gives the detonation adiabat. Conditions in detonation products correspond to the intersection of the same R-M line with the detonation adiabat. In case of a nuclear statistical equilibrium (NSE) $\bar{Y}_d = \bar{Y}_{d,NSE}(T_d, \rho_d)$ is a unique function of temperature and density and $q_d = \bar{Q} \cdot \bar{Y}_{d,NSE} = q_{d,NSE}(T_d, \rho_d)$ is a function of $T_d$ and $\rho_d$ as well. The detonation adiabat with $q = q_{d,NSE}$ is marked on Fig. 1 as $NSE$. The evolution of chemical species variables $\bar{Y}$ with $x$ is described by a steady-state form of (4),

$$\frac{d\bar{Y}}{dx} = \frac{1}{u} \bar{R}(\rho, T, \bar{Y}). \tag{10}$$

Conditions inside the reaction zone are determined by intersection of the R-M line with partial adiabats (8) with $q \neq 0$ determined by (9) (ZND; Fickett & Davis (1979)).

The detonation solution $s1$ - $CJ_{NSE}$ in Fig. 1 is the Chapman-Jouguet (CJ) detonation regime with the minimal possible detonation velocity, $D = D_{CJ}$. An unsupported detonation propagates in this regime if $q$ is a monotonic function of $x$. Overdriven detonations are possible in this case for all velocities $D > D_{CJ}$. This is the case for all detonations at $\rho_0 \leq 2 \times 10^7$ g/cm$^3$ considered in the paper. For higher densities the energy release in $0.5C + 0.5O$ detonations is not monotonic and an unsupported detonation propagates in a pathological regime with some velocity $D^* > D_{CJ}$. Overdriven detonations in this case are possible for $D > D^*$. For a few simulations of high-density detonations presented in this paper we used as initial conditions slightly overdriven detonation solutions with $D$ very close to $D^*$. A detailed explanation of a pathological detonation can be found in Fickett & Davis (1979) and for CO detonations in Khokhlov (1989) and Sharpe (1999).

### 3.2. CO detonations with complete energy release

As mentioned in Sect. 1, explosive burning of CO mixtures proceeds through three consecutive stages - burning of $C^{12}$ in the leading $C^{12} + C^{12}$ reaction, burning of $O^{16}$ and the formation of $Si$-group elements, and $Si$-burning which creates $Fe$-peak elements. The state of NSE is reached at the end of $Si$-burning. As a result, the detonation waves in carbon-oxygen have a multi-layered structure with $C$-burning, $O$-burning, and $Si$-burning layers with increasing thickness, $x_C \ll x_O \ll x_{Si}$, which follow each other. The CJ detonation structure for a $0.5C + 0.5O$ detonation at $\rho_0 = 3 \times 10^6$ g/cm$^3$ is shown in Fig. 2 as an example. The distinct $C$-burning, $O$-burning and $Si$-burning layers are clearly visible. $C$-burning releases $q_C \simeq 0.35$ MeV/nucleon of nuclear energy, at the end of the $O$-burning the total released energy is $q_O \simeq 0.65$ MeV/nucleon, and the total energy released at the end of the reaction zone is $q_{d,NSE} \simeq 0.8$ MeV/nucleon. The detonation products at this density are mostly composed of $Ni$.

In what follows we will characterize detonations in CO mixtures with the overdrive parameter

$$f = \left(\frac{D}{D_{CJ,NSE}}\right)^2 \quad (11)$$

calculated with respect to the CJ detonation velocity of the complete detonation in which burning proceeds to NSE. Pathological detonation occurring at $\rho_0 > 2 \times 10^7$ g/cm$^3$ has an overdrive $f^* = (D^*/D_{CJ,NSE})^2 > 1$. Calculations show that deviation of $f^*$ from one are relatively small. For $\rho_0 = 3 \times 10^7$ g/cm$^3$ $f^* \simeq 1.01$. For higher densities $f^*$ increases to $f^* \simeq 1.07$. Fig. 3 shows an overdriven detonation solution for $\rho_0 = 10^8$ g/cm$^3$ with $f = 1.08$ which is very close to the pathological regime at this density. Again, the distinct $C$-burning, $O$-burning and $Si$-burning layers can be clearly seen. Similar to the previous case, $C$-burning releases $q_C \simeq 0.36$ MeV/nucleon of nuclear energy but due to photo-dissociation of nuclei in $O$-burning and $Si$-burning layers $q_O$ is reduced to $\simeq 0.5$ MeV/nucleon and the overall detonation energy to $q_{d,NSE} \simeq 0.37$ MeV/nucleon, resulting in the non-monotonic energy release. For detonations with non-monotone energy release we find that maximum of $q$ occurs near the maximum of the concentration of $Si$. The energy release in $C$-burning layer is always monotonic so that $C$-detonation will always be not pathological.

Fig. 4 gives $x_C$, $x_O$, and $x_{Si}$ for unsupported $0.5C + 0.5O$ detonations as a function of $\rho_0$; $x_C$ and $x_O$ are defined as half-reaction zone thickness where the concentrations of $C^{12}$ and $O^{16}$ diminish to one-half of their initial values, respectively; $x_{Si}$, is defined as a thickness where $Si$ decreases to one-half of its maximum value inside the reaction zone. The total thickness of the detonation wave, $x_{NSE}$, corresponds to 90 % of the $Ni$ synthesized in the detonation wave. The horizontal dotted line in Fig. 4 shows the characteristic size of an exploding WD, $R_{WD} \sim 10^8$ cm. For $\rho \leq \rho_{NSE} \simeq 10^7$ g/cm$^3$ $x_{NSE} \geq R_{WD}$ which means that burning in a detonation wave will not have time to reach the state of NSE. At $\rho \leq \rho_{Si} \simeq 5 \times 10^6$ g/cm$^3$ the reaction zone of a CJ detonation will consists of $C$-burning layer followed by $O$-burning layer but $Si$-layer will be absent. At $\rho \leq \rho_O \simeq 10^6$ g/cm$^3$ $x_O \geq R_{WD}$ and the reaction zone will consist of $C$-burning layer alone.

### 3.3. Carbon detonations with reduced energy release

Due to a large separation between the reaction scales of $C$, $O$, and $Si$-burning, an incomplete detonation may be considered as an isolated $C$-detonation (or an $O$-detonation) wave with reduced energy release $q_C \simeq 0.35$ MeV/nucleon (or $q_O \simeq 0.65$ MeV/nucleon for $O$-detonation). At densities $\rho < 10^8$ g/cm$^3$ photo-dissociation of nuclei is insignificant and $q_C$ and $q_O$ are practically density-independent. For $C$-detonation the situation is schematically illustrated in Fig. 1. The figure shows a $C$-detonation adiabat with energy release $q_d = q_C$. The R-M line $s3$-$CJ_C$ corresponds to a CJ $C$-detonation with velocity $D_{CJ,C} < D_{CJ,NSE}$.

By analogy with (11) we can also characterize $C$-detonations with the overdrive parameter defined relative to $D_{CJ,C}$,

$$f_C = \left(\frac{D}{D_{CJ,C}}\right)^2 = f \cdot \left(\frac{D_{CJ,NSE}}{D_{CJ,C}}\right)^2 > f. \tag{12}$$

Then $C$-burning layer of a CJ NSE detonation, segment $s1$-$c1$ in Fig. 1, can be viewed as a $C$-detonation overdriven to $f_C = (D_{CJ,NSE}/D_{CJ,C})^2$. All $C$-detonations with velocities between $D_{CJ,C}$ and $D_{CJ,NSE}$, such as $s2$-$c2$, can be considered as overdriven detonations with overdrive parameters $1 < f_C < (D_{CJ,NSE}/D_{CJ,C})^2$. Equivalently, they are underdriven detonations characterized by $f$ ranging from $(D_{CJ,C}/D_{CJ,NSE})^2$ to one. The same detonation can be characterized by using either $f$ or $f_C$ which are related by (12). In this paper we choose to uniformly parametrize all detonations using $f$. The above consideration equally applies to $O$-detonations with the replacement of $q_C$ with $q_O$ and $D_{CJ,C}$ with the corresponding value of $D_{CJ,O}$.

Since $C$ and $O$-detonations have lower energy release, their temporal and spatial scales are larger than those of the NSE detonations. Fig. 5 shows the reaction zone length $x_C$ and $x_O$ for $C$-detonations and $O$-detonations as a function of $f$ for densities $\rho = 10^6$, $3 \times 10^6$, $10^7$ g/cm$^3$. The curves begin on the left at the values of $f$ which correspond to the CJ regimes of $C$ and $O$-detonations. On the right, all curves terminate at $f = 1$ where the $C$ and $O$-detonation becomes a part of a complete CJ NSE detonation moving with velocity $D_{CJ,NSE}$. Spatial scales of $C$ and $O$-detonations increase with decreasing $f$ exponentially. For CJ $C$-detonation $x_C$ exceeds the thickness of the $f = 1$ detonations by a factor $\simeq 30 - 100$.

## 4. One-dimensional propagation of CO detonations

### 4.1. Detonation stability

Numerical simulations of detonation propagation are listed in Table. 1. Details of a numerical approach and test simulations are described in Appendix A. For each simulation the table gives the initial density $\rho_0$, the overdrive $f$ and other parameters of a steady-state detonation. The column seven of the table describes the non-steady behavior of a detonation wave found in the simulations: "S" - stable propagation, "O" and "D" - unstable propagation. In O-type cases we observed several cycles of oscillations of the detonation wave. In D-type cases the detonation rapidly decayed, the leading shock wave decoupled from the reaction zone and the re-ignition of the detonation did not occur over the time of numerical integration. This does not mean that the detonation dies and never re-ignites but that the re-ignition does not occur on the time scale of at least $\simeq L/a_s$, where $L$ is the length of the computational domain and $a_s$ is the sound speed. The re-ignition time for D-type detonations is estimated in Sect. 4.3.

Fig. 6 summarizes the stability properties of the calculated detonations on the $\rho_0$ - $f$ plane. The boundary of detonation stability for NSE detonations in $0.5C+0.5O$ mixtures has been previously calculated in Khokhlov (1993). It was found that the CJ NSE detonations ($f = 1$) are stable for $\rho_0 < 2 \times 10^7$ g/cm$^3$. For higher densities the boundary of stability passes through overdriven detonation regimes with $f > 1$. The stability curve shown in Fig. 6 combines the results of Khokhlov (1993) for $\rho_0 > 2 \times 10^7$ g/cm$^3$ with the results of this paper for $\rho_0 < 2 \times 10^7$ g/cm$^3$. At low densities the curve passes through overdriven regimes of $C$-detonations with $f < 1$. The stability curve crosses the $f = 1$ line at $\rho_0 = 2 \times 10^7$ g/cm$^3$. It must be stressed that at all densities the regimes located at the stability curve are characterized by $f_C > 1$ and are overdriven with respect to the CJ $C$-detonation. The freely propagating $C$-detonations are highly unstable and exhibit the D-type behavior. Overdriven detonations with $f = 1.08$ calculated for high densities $\rho_0 \geq 10^8$ g/cm$^3$ have the structure of the $C$-burning layer which is very close to that of freely propagating pathological detonations. These detonations are highly unstable and exhibit the D-type behavior as well.

Changes in the detonation behavior when $\rho_0$ is fixed and $f$ is decreasing are illustrated in Fig. 7 for $\rho_0 = 3 \times 10^6$ g/cm$^3$. At this density the detonation with $f = 0.76$ is stable whereas the detonation with $f = 0.71$ is unstable, from which we may deduce that the boundary of stability at this density is located between these two values of $f$. The linear interpolation gives $f_s \simeq 0.73$. The detonation with $f = 0.71$ is mildly unstable (Fig. 7a). The initial slow growth of perturbations is followed by regular oscillations with a period $\Pi \simeq 9.5 t_C$ and an amplitude $\simeq 0.1$ of the steady-state post-shock pressure. For the detonation with $f = 0.62$ the instability develops more rapidly. The quasi-periodic oscillations have a much larger period, $\Pi \simeq 50 t_C$, and a much large amplitude $\simeq 4$ of the steady-state post-shock pressure (Fig. 7b). For a detonation with $f = 0.52$ the initial rapid increase of post-shock pressure is followed by a rapid decay of a detonation wave (Fig. 7c). Fig. 7d compares calculations of the $f = 0.55$ detonation made with $\Delta \simeq x_C/30$ and $L = 100 x_C$ (dots) and with $\Delta \simeq x_C/100$ and $L = 640 x_C$ (solid line). The comparison shows that the behavior is resolution-independent and not influenced by the boundary conditions.

### 4.2. Oscillation cycle

Figs. 8, 9 and 10 show the distributions of pressure, temperature, and carbon mass fraction inside the reaction zone at different moments of time during the third cycle of oscillations (Fig. 7b) for the detonation $\rho_0 = 3 \times 10^6$ g/cm$^3$ and $f = 0.62$. During the minimum of post-shock pressure, times $t_1$ to $t_3$, the leading shock is weakened and the post-shock temperature is smaller than the corresponding steady-state post-shock temperature. As a result the reaction zone is approximately ten times wider than $x_C$.

An important consequence of the widening of the reaction zone is the decrease of the gradients of temperature and reactivity of matter behind the leading shock. Flattening of the gradient of reactivity causes the development of a supersonic spontaneous reaction wave behind the leading shock between times $t_3$ and $t_4$. The emergence and amplification of the spontaneous wave is clearly visible in Fig. 8 at $t_4$ and $t_5$ as a growing peak of pressure propagating to the right. Between $t_5$ and $t_6$ the forward part of the spontaneous wave steepens into a shock and the spontaneous wave transforms into a secondary detonation wave which propagates through the shocked and compressed un-reacted matter towards the leading shock.

Between $t_6$ and $t_7$ the secondary detonation overcomes and interacts with the leading shock and passes into the cold unburned matter where it becomes the main detonation wave. The interaction of the secondary detonation with the leading shock generates a rarefaction wave which travels back into the burned material. The rarefaction leads to a gradual weakening of the leading shock and widening of the reaction zone ($t_8$ to $t_{12}$). By the time $t_{12}$ the leading shock is weakened to a degree that the distribution of physical parameters behind the shock resembles the distribution which existed in the beginning of the oscillation cycle at time $t_1$. At this moment the new oscillation cycle begins.

Generation of a spontaneous wave and the subsequent transition to a detonation in compressed matter occurs via the Zeldovich gradient mechanism (Zeldovich et al. 1970;

Lee & Moen 1980) in which the pressure wave is amplified by the energy release source which travels with the spontaneous velocity $D_{sp} = \left(\frac{d\tau(T(x))}{dx}\right)^{-1}$, where $\tau(T)$ is the energy release reaction timescale. For an efficient amplification of a pressure wave, $D_{sp}$ must be comparable to or exceed the local speed of sound, and the spontaneous reaction source must act for a sufficient period of time. The diminishing of the temperature gradient caused by the weakening of the leading shock favors both these requirements.

Since $\tau$ is an exponential function of $T$, the size of the reaction zone during the minimum of the detonation cycle depends exponentially on $f$. For $f = 0.71$, the increase of the reaction zone at the minimum of oscillations is much less pronounced than for $f = 0.62$. As a result $D_{sp}$ becomes smaller and the generated pressure wave is weak and does not transition to a secondary detonation before it reaches the leading shock. In this case we observe a low amplitude, nearly sinusoidal oscillating behavior of a detonation wave. With decreasing $f$ the length of the reaction zone at minimum increases dramatically. In the D-case $f = 0.55$ the reaction zone rapidly separates from the leading shock and the distance between the shock and the reacted matter becomes much larger than the length of the computational domain $L$. In this case we observe the decay of the detonation wave but the re-ignition and formation of a spontaneous wave is significantly delayed and is not observed.

### 4.3. Re-ignition time

We now proceed to estimate a characteristic time-scale of re-ignition for D-type detonations. Fig. 11 illustrates the D-type behavior of a detonation with $\rho_0 = 3 \times 10^6$ g/cm$^3$ and $f = 0.55$. In this case the initial perturbations grow and then the reaction zone rapidly separates from the leading shock. The post-shock pressure decreases from the initial values $P \simeq 13P_0$ (time $t_1$) to $P \simeq 8P_0$ at $t_7$, and temperature from $T_s \simeq 2.6 \times 10^9$K to $T_s \simeq 1.8 \times 10^9$K. By the time $t_7$ the detonation pressure spike disappears and the pressure in the shocked material becomes constant and approximately equal to the pressure $P_d$ of the corresponding steady state detonation.

The re-ignition time can thus be estimated as follows. For a given $\rho_0$ and $f$ we first calculate the detonation pressure $P_d$ assuming $q_d = q_C$. Next, we calculate the post-shock temperature $T_s$ and post-shock density $\rho_s$ of a shock with the post-shock pressure $P_s = P_d$. Finally, using the values of $T_s$, $\rho_s$ and the initial carbon composition $Y_{C,0}$ we calculate the timescale $\tau_C(T_s, \rho_s, Y_{C,0})$ for the leading $C^{12} + C^{12}$ reaction. This time serves as an estimate of the reignition time.

We calculate $\tau_C$ as follows. The kinetic of the $C^{12} + C^{12}$ reaction may be approximated

as
$$\frac{dY_C}{dt} = -R_C(\rho, T, Y_C) = -A\rho Y_C^2 \, e^{-Q/T_a^{1/3}}, \tag{13}$$

where $Q \simeq 84.165 \times 10^3 \text{K}^{1/3}$, $T_a = T/(1 + 0.067 \times 10^{-9} T)$, and $A(T)$ is a known weak (non-exponential) function of temperature (Fowler et al. 1975).

In what follows the subscript $i$ indicates initial physical values at the beginning of the reaction. The e-folding time of $C^{12} + C^{12}$ reaction is

$$\tau_{C,i}^{-1} = -\frac{1}{Y_{C,i}} \frac{dY_{C,i}}{dt} = \left(\frac{R_C}{Y_C}\right)_i \tag{14}$$

if we assume that the reaction occurs at constant $T = T_i$ and $\rho = \rho_i$. To evaluate the actual time-scale $\tau_C$ we must take into account self-acceleration of the $C^{12} + C^{12}$ reaction caused by increase of $T$.

The evolution of $T$ is governed by

$$\frac{dT}{dt} = \left(\frac{\partial T}{\partial \epsilon}\right) \left(\frac{q_C}{Y_{C,i}}\right) R_C(T, \rho, Y_C), \tag{15}$$

where $q_C$ is the energy release in $C^{12}+C^{12}$ reaction per unit mass of CO mixture with initial carbon composition $Y_{C,i}$. Using Frank-Kamenetskii (1967) (FK) approximation we assume that all values in (15) can be evaluated at constant initial conditions except the exponent in $R_C(T)$. In FK approximation we can re-write (15) as

$$\frac{dT}{dt} \simeq \left(\frac{T_i}{\tau_{T,i}}\right) \exp\left(\frac{Q}{T_{a,i}^{1/3}} - \frac{Q}{T_a^{1/3}}\right) \simeq \left(\frac{T_i}{\tau_{T,i}}\right) \exp\left(\frac{1}{\Theta}\left(\frac{T}{T_i} - 1\right)\right) \tag{16}$$

where

$$\tau_{T,i} = \left(\frac{\epsilon_i}{q_C}\right) \left(\frac{\partial \ln \epsilon_i}{\partial \ln T_i}\right) \tau_{C,i} \tag{17}$$

is the temperature e-folding time evaluated at constant initial conditions and

$$\Theta = \left(\frac{3T_{a,i}^{1/3}}{Q}\right) \cdot \left(\frac{T_i}{T_{ai}}\right) \tag{18}$$

is the Frank-Kamenetskii factor. Integration of (16) gives

$$\tau_C \simeq \tau_{T,i} \int_{T_i}^{\infty} \exp\left(\Theta^{-1}\left(\frac{T}{T_i} - 1\right)\right) d\left(\frac{T}{T_i}\right) = \Theta \cdot \tau_{T,i}. \tag{19}$$

Formally, the upper limit in the integral (19) should be equal to some final temperature $T_f > T_i$ reached at the end of burning. $T_f$ may be replaced with $\infty$ since the integral in

(19) is mostly accumulated at temperatures $T \simeq T_i$. FK factor $\Theta$ in (19) characterizes shortening of the reaction timescale due to self-acceleration of burning. We note that in the FK approximation

$$\left(\frac{\partial \ln R_i}{\partial \ln T_i}\right) \simeq \frac{1}{\Theta} \qquad (20)$$

so that $\tau_C$ may also be written as

$$\tau_C(T_i) \simeq \left(\frac{\epsilon_i}{q_C}\right)\left(\frac{\partial \ln \epsilon_i}{\partial \ln T_i}\right)\left(\frac{\partial \ln T_i}{\partial \ln R_i}\right)\tau_{C,i}. \qquad (21)$$

Fig. 12 shows temporal and spatial re-ignition scales as a function of $\rho_0$ which were estimated for the borderline D-cases (see Table 1 and dashed line in Fig. 6). Both scales increase exponentially with decreasing $\rho_0$. At densities less than $\simeq 10^6$ g/cm$^3$ the time scale becomes greater than the explosion time of a white dwarf ($\simeq 1$ sec) and the spatial scale, estimated by multiplying the velocity of the leading shock of the detonation with the reignition time, becomes greater than the scale of the exploding white dwarf. At these densities the one-dimensional propagation of a $C$-detonation will be impossible.

### 4.4. Stability criterion

In general, the instability is a consequence of a positive feedback between the hydrodynamical perturbations and the temperature-dependent energy release inside the detonation wave. The feedback loop consists of three parts. First, the temperature perturbation leads to the increase of the reaction rate. This effect depends on the sensitivity of $R$ to $T$. Second, the increase of $R$ leads to the increase of the energy generation rate. This effect is proportional to the amount of nuclear energy which is released by burning. Third, the increase in the energy generation leads to a faster rise of $T$. The latter effects depends on specific heat of matter. For illustration, consider a thermally isolated, uniform fluid element of a carbon-oxygen mixture at constant volume. The evolution of $T$ in the element is described by (15). To compare the feedback loop for fluid elements with different initial temperatures and densities we scale $t$ to the time $\tau_{C,i}$ which is required to burn half of initial carbon at constant initial conditions (see (14)). Then in FK approximation (15) becomes

$$\frac{dT}{d\tilde{t}} = T_i \left(\frac{\partial \ln T_i}{\partial \ln \epsilon_i}\right)\left(\frac{q_C}{\epsilon_i}\right)\frac{R_C(T, \rho_i, Y_{C,i})}{R_C(T_i, \rho_i, Y_{C,i})}, \qquad (22)$$

where $\tilde{t} = t/\tau_{C,i}$.

Consider now two identical fluid elements, one with initial temperature $T_i$ and another one with perturbed initial temperature $T_i + \Delta T$, $\Delta T \ll T_i$. The evolution of $\Delta T$ with $\tilde{t}$

is given by
$$\frac{d\Delta T}{d\tilde{t}} = \frac{d(T+\Delta T)}{d\tilde{t}} - \frac{dT}{d\tilde{t}} \simeq \left(\frac{\partial \ln T_i}{\partial \ln \epsilon_i}\right)\left(\frac{q_C}{\epsilon_i}\right)\left(\frac{\partial \ln R_i}{\partial \ln T_i}\right)\Delta T. \qquad (23)$$

Hence $\Delta T(\tilde{t}) \propto \exp(\beta \tilde{t})$ where the increment
$$\beta = \beta(T_i, \rho_i, Y_{C,i}) = \left(\frac{\partial \ln R_i}{\partial \ln T_i}\right)\left(\frac{q_C}{\epsilon_i}\right)\left(\frac{\partial \ln T_i}{\partial \ln \epsilon_i}\right). \qquad (24)$$

We see that the increment is the product of three dimensionless factors. First factor in (24) describes sensitivity of $R$ to temperature, second factor characterizes the relative energy content of the fuel, and third takes into account specific heat of matter.

Table 1 gives $\beta$ and individual multipliers in (24) for computed detonations. Inspection of the table shows that variations of $\beta$ with $\rho_0$ and $f$ are mostly caused by variations of the second and third multipliers in (24). For example, for $\rho = 10^7$ g/cm$^3$ the value of $\beta$ varies by a factor of $\simeq 2.90$ for $f$ varying from 1 to 0.61. $q_C/\epsilon_s$ and $(\partial \ln T/\partial \ln \epsilon)$ vary by a factor $\simeq 1.6$ whereas $(\partial \ln R/\partial \ln T) = \Theta^{-1}$ varies by a factor of only $\simeq 1.1$. The situation is similar for other densities, as well. The reason for large variations of the second and the third multipliers is related to the dependence of internal energy per unit mass and specific heat of degenerate matter on $\rho$. On the other hand, $\Theta^{-1}$ is density-independent (screening corrections at post-shock temperatures are small and can be neglected) and is a weak function of temperature, $\Theta^{-1} \sim T^{-1/3}$. As a result, relative variations of $\Theta^{-1}$ are small and this factor does not contribute to the variations of $\beta$ significantly.

We note that the relative importance of the factors controlling the instability of a detonation is different in $C$-detonations in supernovae and in terrestrial detonations in reactive gases. The equation of state for terrestrial gases (for example Burcat & Ruscic 2005) is such that $\epsilon_i$ is usually independent of $\rho$ and increases with $T$ linearly or faster (e.g. Burcat & Ruscic 2005). Specific heat in reactive gases is also density-independent and may be a constant or a weak function of $T$. On the other hand, Arrhenius type reaction rates of chemical reactions are $R \sim \exp(-Q/T)$ and thus $\Theta^{-1} = T/Q$ is a stronger function of $T$ compared to $\Theta^{-1}$ for $C$-detonations. As a result, the sensitivity of chemical reaction rates on $T$ play a significant role for terrestrial detonations.

Close inspection of Table 1 shows that the transition from stable to unstable behavior of a $C$-detonation with decreasing $f$ wave typically occurs for $\beta \simeq 9-10$, virtually independent of $\rho_0$. The borderline between O-type and D-type detonations corresponds to $\beta \simeq 14-15$. Of course, the correlation is not perfect. Perfect correlation should not be expected since the detonation wave is a distributed system in which physical conditions vary inside the reaction zone. Communication of information from one point to another involves interaction of many fluid elements and occurs with a finite speed. Derivation of (24) is done within a simple one-zone model and does not take any of this into account. Nevertheless, it seems that the derivation catches a significant part of the feedback.

Finally, the feedback analysis helps to understand why detonation in degenerate carbon-oxygen tends to become more unstable at higher densities. The leading part of a complete CO detonation can be viewed as a $C$-detonation overdriven to $f \geq 1$. The feedback loop consideration applies to this detonation as well. With increasing $\rho_0$ specific heat of degenerate matter decreases drastically and as a result $(\partial \ln T/\partial \ln \epsilon)$ in (24) increases by a factor of almost $\simeq 10$ (see Table 1). This leads to a noticeable increase of the strength of the feedback loop. As a result, greater overdrive is needed for stabilizing the detonation and the stability boundary passes through the overdriven detonation regimes (Sect. 4.1 and Fig. 6).

## 5. Discussion and Conclusions

In this paper we studied basic properties and one-dimensional stability of incomplete carbon-oxygen detonations in low density environment of SN Ia. Calculations of the steady-state structure of $C$-detonations show that a Chapman-Jouguet $C$-detonation has a velocity $D_{C,CJ}$ which is a factor of $\simeq 20 - 25\%$ smaller than the CJ velocity of a complete NSE detonation $D_{CJ,NSE}$. This decrease translates into the decrease of the post-shock temperature by a factor $\simeq 0.7$ (Table 1), and the corresponding increase of the half-reaction thickness $x_C$ by a factor $\simeq 30 - 100$ (Fig. 5). The half-reaction time scale $t_C \simeq x_C/D$ increases by approximately the same factor. While the increase is appreciable, at all relevant densities $x_C$ still remains relatively small compared to the the characteristic scale of a WD, $\simeq 10^8$ cm.

Decreased energy release in $C$-detonations has a major effect on the detonation stability. In contrast to complete detonations which are one-dimensionally stable at low densities, $C$-detonations are highly one-dimensionally unstable. The boundary of stability for $C$-detonations (Fig. 6) passes through detonation regimes which are overdriven with respect to CJ $C$-detonation. Near the boundary of stability unstable detonations exhibit low-amplitude pulsations and have a continuous distribution of physical parameters inside the reaction zone. With decreasing overdrive the detonation becomes highly unstable and begins to propagate by periodically generating a spontaneous reaction wave inside the primary reaction zone. The spontaneous wave transitions to a secondary detonation which overcomes and accelerates the leading detonation shock. The subsequent decay and the separation of the leading shock from the reaction zone gives rise to a new spontaneous wave and starts the next oscillation cycle.

Transition from stable to unstable behavior, the level of instability, and the length of the detonation cycle depend mainly on specific heat and internal energy of degenerate matter behind the leading shock. The sensitivity of reaction rates to post-shock temperature is less important. This is in contrast with idealized Arrhenius detonations whose stability is usually highly dependent on the sensitivity of the reaction rate to temperature (e.g. Lee & Stewart 1990). Stability and detonation cycle of a $C$-detonation in SN Ia cannot be understood without taking effects of the equation of state of degenerate matter into account.

Thickness of a steady-state one-dimensional detonation wave (Fig. 5) can be used to estimate the minimal resolution required for a correct description of the detonation structure and detonation stability. The absolute minimum of numerical resolution can be no less than $n \simeq 10 - 30$ computational cells per $x_C$. The minimal spatial resolution $\Delta r$ depends on background conditions in unburned matter and can be estimated for a given density and overdrive by taking $x_C$ from Fig. 5 and dividing it by $n$. Even for the most favorable case of a CJ $C$-detonation at $\rho_0 = 10^6$ g/cm$^3$ (the upper left end of the solid line for $\lg \rho_0 = 6$ in Fig. 5) and $n = 10$ the required resolution $\delta r \simeq 10^4$ cm is significantly less than the resolution of current global three-dimensional simulations of SN Ia. For one-dimensional lagrangian simulations we can estimate the relative mass resolution $\delta q = \delta M / M_{WD}$ as

$$\delta q \simeq \frac{4\pi \rho_0 R^2 x_C}{n M_{WD}} \simeq 10^{-5} \left(\frac{R}{5 \times 10^8 cm}\right)^2 \left(\frac{x_C}{10^4 cm}\right) \left(\frac{M_\odot}{M_{WD}}\right) \left(\frac{10}{n}\right) \left(\frac{\rho_0}{10^6 g/cm^3}\right), \quad (25)$$

where $M$ is the lagrangian mass coordinate and $R$ is the radius of a spherical layer with density $\rho_0$ at the moment of time when the detonation is passing through the layer. For $\rho_0 = 10^6$ g/cm$^3$, $R \simeq 5 \times 10^8$ cm (see below) and $n = 10$ we obtain $\delta q \simeq 10^{-5}$.

The period and spatial scale of the detonation cycle of a $C$-detonation increase with decreasing density. For $\rho \leq 10^6$ g/cm$^3$ the period becomes larger than the explosion time scale, $\sim 1$ sec, of a white dwarf, which suggests that the influence of the instability on propagation and quenching of a $C$-detonation wave could be significant at these densities. This may alter the composition of the outer layers. In particular, quenching at higher densities may leave more unburned carbon. From the observational point of view it would be important to estimate the velocity range of a SN Ia envelope which may be affected. This depends, of course, on a particular explosion model. For illustration, we give estimates for two delayed detonation (DD) models of Höflich et al. (2002). For a normal supernova we take the DD model SP0022.20 with total mass $M_{WD} = 1.346 M_\odot$ and $Ni^{56}$ mass $M_{Ni} = 0.54 M_\odot$. In this model the detonation wave, which propagated outward through the WD already pre-expanded by deflagration, passed through the layer with density $\rho_0 = 10^6$ g/cm$^3$ when the layer was located at the mass coordinate $M = 1.300 M_\odot$ and radius $R \simeq 5 \times 10^8$ cm. At this moment of time the layers with density $\rho_0 \leq 5 \times 10^5$ g/cm$^3$ had radii $R \geq 5.6 \times 10^8$ cm. From this we estimate the characteristic scale of density variation at these densities as $L \simeq \left(\frac{d \ln \rho}{dr}\right)^{-1} \simeq 5 \times 10^7$ cm. The corresponding time-scale of density variation can be estimated by dividing $L$ by the difference in the expansion velocities of the layers $\simeq 10^8$ cm/sec which gives the characteristic time-scale $\simeq 0.5$ sec. The numbers are consistent with the values of $10^8$ cm and 1 sec used in our general analysis in Sections 4 and 3, and in this section above. After the explosion and acceleration to free expansion, the layer with $\rho_0 = 10^6$ g/cm$^3$ was accelerated to $V \simeq 23,000$ km at infinity. From this we conclude that the one-dimensional instability in a normal delayed detonation explosion may affect the outer $0.04 M_\odot$ or 3% of a SN Ia envelope with velocities $V \geq 23,000$ km/sec at infinity. For a sub-luminous DD model 5P0Z22.8 similar analysis shows that the instability may affect outer $0.08 M_\odot$ or 6% of the envelope with velocities $V \geq 14,000$ km/sec. Thus, the instability may affect the spectra and the light curve of these supernovae before and near maximum light. We note that the mass resolution of the models was $\delta q \simeq 10^{-3}$, i.e., two orders of magnitude greater that the resolution required for detecting the detonation instability.

In conclusion we stress that this study is one-dimensional and we caution that the above estimates are preliminary. In addition to a strong one-dimensional instability the detonation will be affected by transversal instability (e.g. Gamezo et al. 1999), curvature of the detonation front (e.g. Sharpe 2001), and spatial variations of the background density. Transverse instability is caused by the same feedback between the energy generation and temperature fluctuations which is responsible for the one-dimensional longitudinal instability. Other effects depend on the strength of the feedback, as well. Thus, incomplete $C$-detonations should be expected to be more unstable and more sensitive to the above-mentioned effects than the complete detonations. Investigation of the multi-dimensional behavior of $C$-detonations will be reported in the next paper (in preparation).

The work was carried out within the NSF project Collaborative research: Three-Dimensional Simulations of Type Ia Supernovae: Constraining Models with Observations. The work was supported by the NSF grant AST-0709181 (A.K.), the Spanish Ministry for Education Mobility Programme within the framework of the National Plan I+D+I 2008-3011, and Spanish Ministry for Science an Innovation project AYA2008-04211-C02-02 (I.D.). The authors are grateful to the members of the collaboration Ed Baron, Peter Hoeflich, Kevin Krisciunas, and Lifan Wang for important discussions. Peter Hoeflich provided detailed numerical data about his delayed detonation models.

Table 1: Properties of NSE and C-detonations in $0.5C^{12} + 0.5O^{16}$ mixtures.

| $\rho_0$[a] | $f$[b] | $\rho_s/\rho_0$ | $P_s/P_0$ | $T_s$[a] | $D$[a] | Stab | $\Theta^{-1}$ | $\left(\frac{\partial \ln T}{\partial \ln \epsilon}\right)_\rho$ | $\frac{q_C}{\epsilon_s}$ | $\beta$ |
|---|---|---|---|---|---|---|---|---|---|---|
| 1.0 $10^6$ | 1.00 | 5.16 | 39.4 | 3.17 | 1.17 | S | 19.6 | 0.43 | 0.65 | 6.14 |
| | 0.79 | 4.84 | 30.5 | 2.81 | 1.04 | S | 20.4 | 0.52 | 0.77 | 8.21 |
| | 0.71 | 4.69 | 27.3 | 2.65 | 0.98 | O | 20.8 | 0.57 | 0.86 | 10.1 |
| | 0.58 | 4.43 | 22.2 | 2.35 | 0.89 | O | 21.7 | 0.67 | 1.03 | 15.0 |
| | 0.55 | 4.35 | 21.0 | 2.27 | 0.87 | D | 21.9 | 0.70 | 1.08 | 16.6 |
| 3.0 $10^6$ | 1.00 | 4.80 | 24.3 | 3.85 | 1.21 | S | 18.2 | 0.55 | 0.52 | 5.17 |
| | 0.77 | 4.47 | 19.1 | 3.35 | 1.07 | S | 19.2 | 0.71 | 0.67 | 9.15 |
| | 0.71 | 4.33 | 17.1 | 3.12 | 1.02 | O | 19.7 | 0.79 | 0.74 | 11.4 |
| | 0.62 | 4.13 | 15.0 | 2.86 | 0.96 | O | 20.3 | 0.88 | 0.82 | 14.6 |
| | 0.58 | 4.03 | 14.0 | 2.72 | 0.92 | D | 20.6 | 0.94 | 0.87 | 16.8 |
| | 0.55 | 3.97 | 13.2 | 2.60 | 0.90 | D | 20.9 | 0.99 | 0.91 | 18.9 |
| 1.0 $10^7$ | 1.00 | 4.24 | 14.0 | 4.42 | 1.21 | S | 17.6 | 0.90 | 0.48 | 7.62 |
| | 0.86 | 4.01 | 11.9 | 3.94 | 1.12 | S | 18.2 | 1.06 | 0.54 | 10.5 |
| | 0.77 | 3.85 | 10.7 | 3.62 | 1.06 | O | 18.7 | 1.19 | 0.59 | 13.2 |
| | 0.74 | 3.77 | 10.1 | 3.45 | 1.03 | D | 19.1 | 1.27 | 0.62 | 14.9 |
| | 0.69 | 3.67 | 9.49 | 3.27 | 1.00 | D | 19.4 | 1.36 | 0.64 | 17.0 |
| | 0.61 | 3.49 | 8.40 | 2.93 | 0.95 | D | 20.1 | 1.56 | 0.70 | 22.0 |
| 3.0 $10^7$ | 1.02 | 3.57 | 8.20 | 4.39 | 1.16 | O | 17.6 | 1.63 | 0.43 | 12.4 |
| | 0.94 | 3.45 | 7.64 | 4.12 | 1.13 | D | 18.0 | 1.78 | 0.45 | 14.4 |
| | 0.94 | 3.42 | 7.49 | 4.04 | 1.12 | D | 18.1 | 1.86 | 0.46 | 15.1 |
| 1.0 $10^8$ | 1.08 | 2.95 | 5.30 | 4.36 | 1.19 | D | 17.6 | 3.10 | 0.36 | 19.6 |
| 3.0 $10^8$ | 1.08 | 2.49 | 3.82 | 4.16 | 1.23 | D | 17.9 | 6.00 | 0.32 | 34.4 |

---

[a]$\rho_0$ is in g/cm$^3$; $T_s$ is in $10^9$K; $D$ is in $10^9$cm/s.

[b]Detonations with $f \geq 1$ are calculated assuming NSE in products of burning (Sect. 3.2). Detonations with $f < 1$ are C-detonations (Sect. 3.3).

## A. Numerical method

### A.1. Numerical approach

The equations (1) - (4) are integrated using an adaptive mesh refinement (AMR) reactive flow fluid dynamic code ALLA (Khokhlov 1998). Reaction terms are coupled to fluid dynamics by time-step splitting. During the hydrodynamical sub-step the Euler equations are integrated with $\bar{R} = \dot{q} = 0$ using an explicit, directional-split, second-order accurate, Godunov-type conservative scheme with a Riemann solver (Colella & Glaz 1985) and a monotone VanLeer reconstruction (van Leer 1979). Hydrodynamical time step is selected using the Courant number = 0.9. During the reaction sub-step the kinetic equations together with the equation of energy conservation, differential equations

$$\frac{d\bar{Y}}{dt} = \bar{R}, \quad \frac{d\epsilon}{dt} = \dot{q}, \tag{A1}$$

are integrated in each cell using a stiff solver with adjustable sub-cycling to keep the accuracy of integration of energy and composition to better than $10^{-3}$.

AMR is carried out at the level of individual cells. Size of computational cells at refinement level $l_{min} < l < l_{max}$ is $\Delta(l) = L \cdot 2^{-l}$; where $l_{max}$ and $l_{min}$ are maximum and minimum levels of refinement, respectively. During the simulation the mesh is refined around the shocks, contact discontinuities, and in regions of large gradients of density, pressure, and mole fractions of $H^4$, $C^{12}$, $O^{16}$, $Ne^{20}$, $Mg^{24}$ and $Si^{28}$. Detected shocks and discontinuities are always refined to maximum level of resolution. Refinement based on chemical composition is used for species with concentrations $Y_i > 10^{-3}$. Further details of the refinement procedure can be found in Khokhlov (1998).

Calculations are performed in a reference frame moving with a constant velocity of a steady-state detonation $D$ in order to keep the detonation on the mesh as long as possible. A steady state solution corresponding to a detonation velocity $D$ is mapped onto a computational domain of size $L$. A constant supersonic inflow with $u = -D$, $\rho = \rho_0$, $P = P_0$, and $\bar{Y} = \bar{Y}_0$ is imposed on the right boundary $x = L$. Zero-gradient boundary conditions are imposed at the left boundary $x = 0$. Calculations were performed with numerical resolution of $\simeq 30$ cells per half-reaction length, $\Delta \simeq x_C/30$, and $L \simeq 100 x_C$ and then repeated with two times lower numerical resolution and, when necessary, with up to four times higher numerical resolution and $L \simeq (300 - 1000) x_C$ in order to confirm the observed detonation behavior.

## A.2. Test simulations

Linear stability and one-dimensional propagation of idealized detonations with the one-step Arrhenius kinetics and $\gamma$-constant equation of state,

$$\frac{dY}{dt} = -Ye^{-\frac{Q}{T}}, \quad \frac{d\epsilon}{dt} = -q\frac{dY}{dt}, \quad P = (\gamma - 1)\rho\epsilon, \quad T = \frac{P}{\rho}, \tag{A2}$$

where $Q$ is the activation energy, $q$ is the total energy release, and $\gamma = 1.2$ have been extensively studied in the past (Erpenbeck 1964; Lee & Stewart 1990; He & Lee 1995; Williams et al. 1996; Sharpe & Falle 2000 and references therein). We carried out a number of test simulations in order to verify that the code can reproduce the stability properties of these detonations correctly.

Fig. 13a shows pressure $P_s$ as a function of $t$ for an overdriven detonation $Q = q = 50$ and $f = 1.6$. For this case the linear analysis predicts a single unstable mode $\propto e^{\sigma t}$ with the increment $\sigma = \sigma_R + i\sigma_I = 0.112 \pm i0.789$; the corresponding period of oscillations is $\Pi = 2\pi/\sigma_I = 7.963$ (Erpenbeck 1964; Lee & Stewart 1990). The simulations show initial growth of perturbations with $\sigma_R \simeq 0.1$. The amplitude of perturbations saturates due to non-linear effects at the level of $\simeq 0.8$ of the steady-state post-shock pressure. The numerical period of oscillations $\Pi \simeq 8.42$ is larger than the linear prediction. This is consistent with previous numerical simulations of Fickett & Davis (1979); He & Lee (1995); Williams et al. (1996); Sharpe & Falle (2000). The increase of $\Pi$ can be attributed to the non-linearity of oscillations which manifests itself in a large amplitude of pulsations and in a pronounced asymmetry in the shape of $P_s(t)$. According to the linear analysis the transition from stable to unstable behavior of the detonations with $Q = q = 50$ must occur at $f = 1.731$ (Lee & Stewart 1990). In numerical simulations we find the transition at $f \simeq 1.75$ in a reasonable 1% agreement with the linear analysis.

Figs. 13(b-d) show results of simulations for CJ detonations with $f = 1$, $q = 50$ and varying $Q$. The linear stability analysis predicts the boundary of stability at $Q \simeq 25.26$ (He & Lee 1995). For $Q = 25$ the authors predict the stable root $\sigma = -0.012 \pm i0.494$. Our simulations show a stable propagation for a detonation with $Q = 24$ (Fig. 13b). For $Q = 25$ the simulations (Fig. 13c) show quasi-periodic slowly decaying oscillations with $\sigma \simeq -0.011 + i0.49$ in good agreement with the linear predictions. For $Q = 27$, interpolation in Table II of He & Lee (1995) gives $\sigma_I \simeq 0.470$. The simulations Fig. 13d give similar $\sigma_I \simeq 0.47$ in good agreement with the linear analysis and non-linear calculations of He & Lee (1995) (their Figure 10b). The instability of a detonation tends to increase with decreasing $f$. In agreement with He & Lee (1995); Williams et al. (1996); Sharpe & Falle (2000) we found that the propagation of the above detonations becomes highly irregular when $f$ approaches $f = 1$.

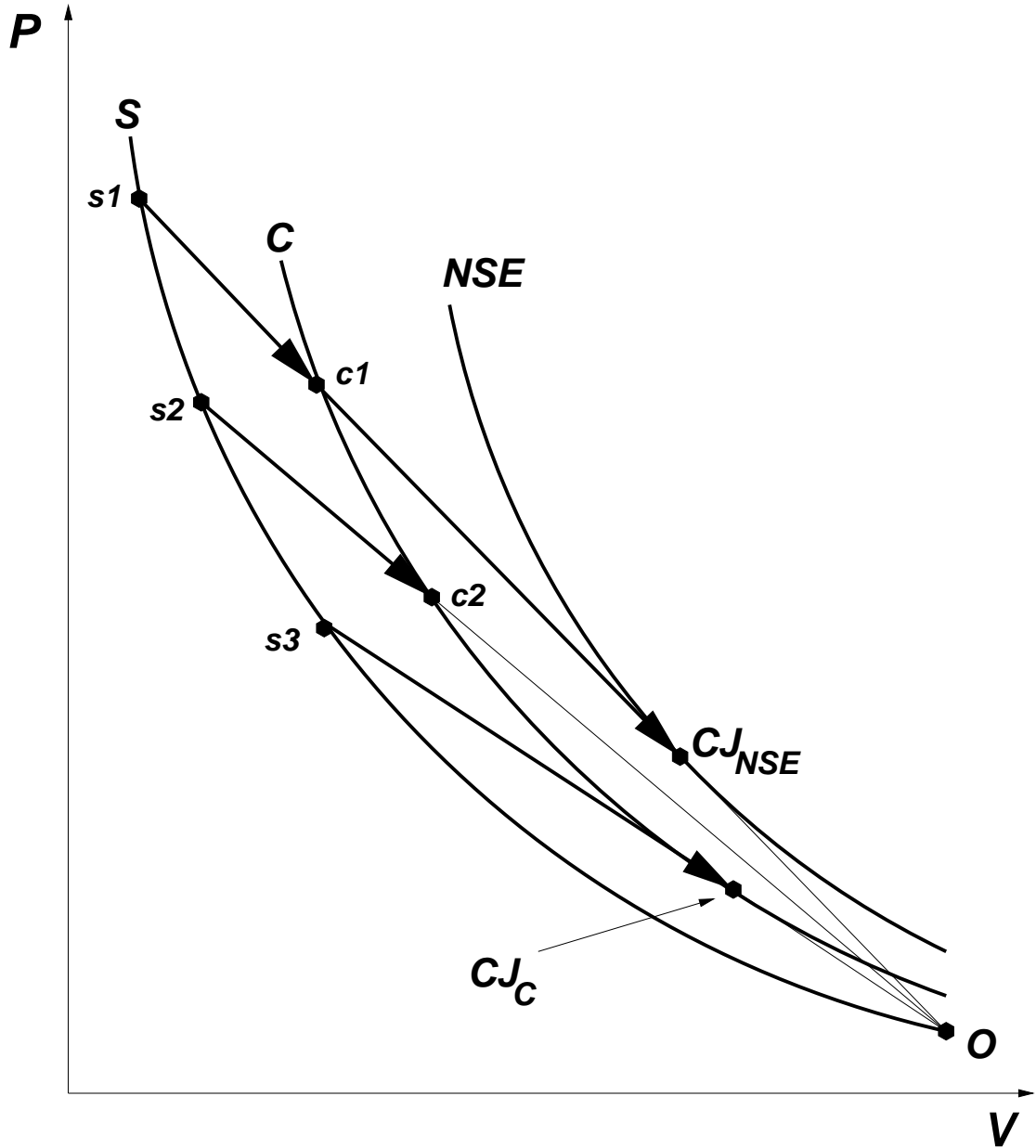

Fig. 1.— Schematic representation of NSE and incomplete detonation waves on a pressure $P$ - specific volume $V = 1/\rho$ diagram. O - initial state of matter ahead of the shock. S - shock adiabat, NSE - detonation adiabat, C - partial detonation adiabat corresponding to C-burning with reduced energy release $q_d = q_C < q_{NSE}$. $s1$ - $\text{CJ}_{NSE}$ - Chapman-Jouguet NSE detonation; $s1$ - post-shock conditions for the CJ NSE detonation; $c1$ - conditions at the end of the C-burning layer of CJ NSE detonation. $s3$ - $\text{CJ}_C$ - Chapman-Jouguet C-detonation, $s3$ - post-shock conditions for the CJ C-detonation. $s2$ - $c2$ - a C-detonation overdriven with respect to the CJ C-detonation and underdriven with respect to the CJ NSE detonation.

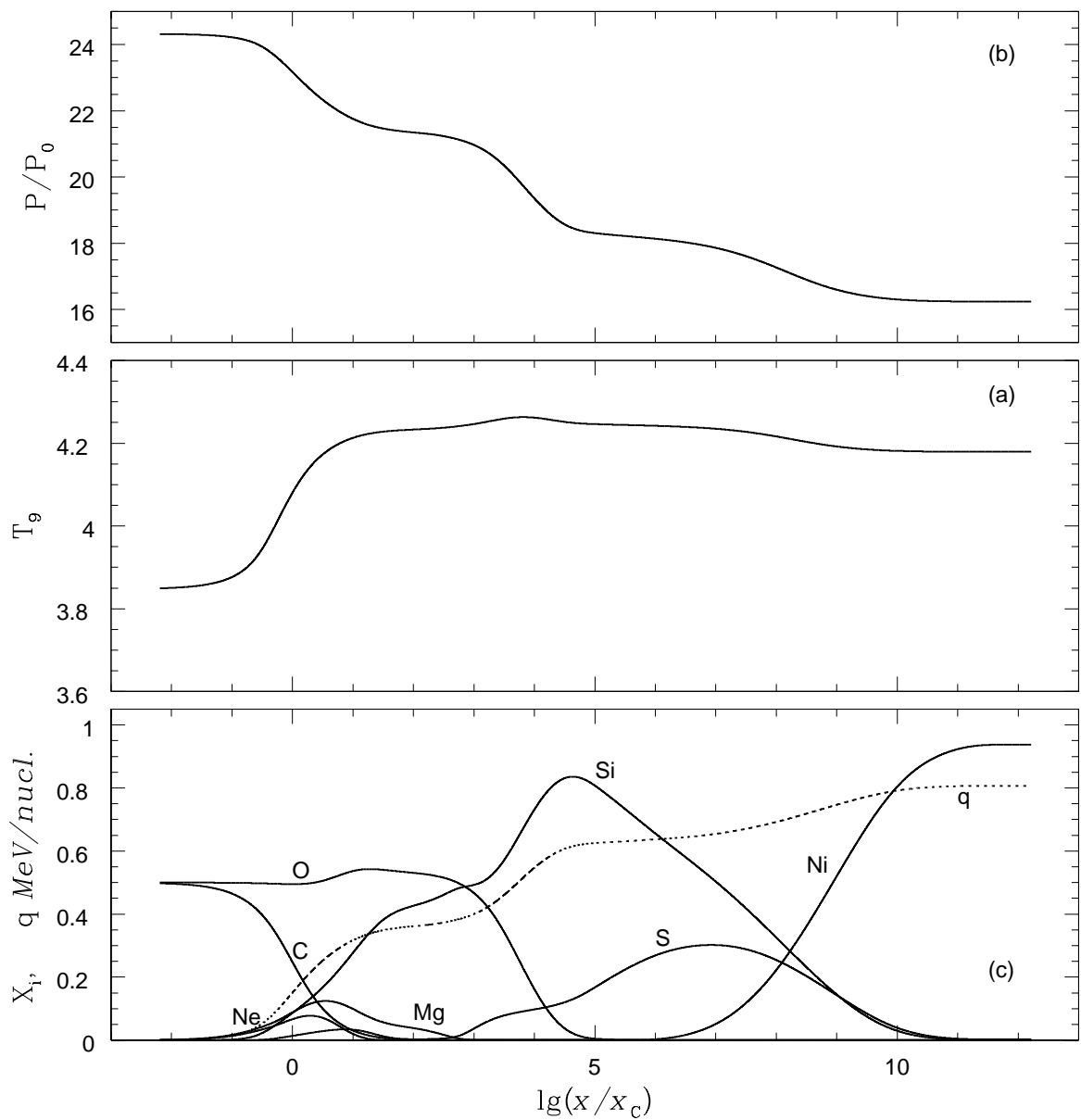

Fig. 2.— A Chapman-Jouguet NSE detonation structure for $0.5C+0.5O$ mixture at $\rho_0 = 3 \times 10^6$ g/cm$^3$ and $f = 1$. Top - pressure, middle - temperature in $10^9$K, bottom - mass fractions of nuclei (solid curves) and nuclear energy release (dashed line). Distance is normalized to a half-reaction length of a C-burning layer. Shock wave is on the left (at $-\infty$ in logarithmic coordinates).

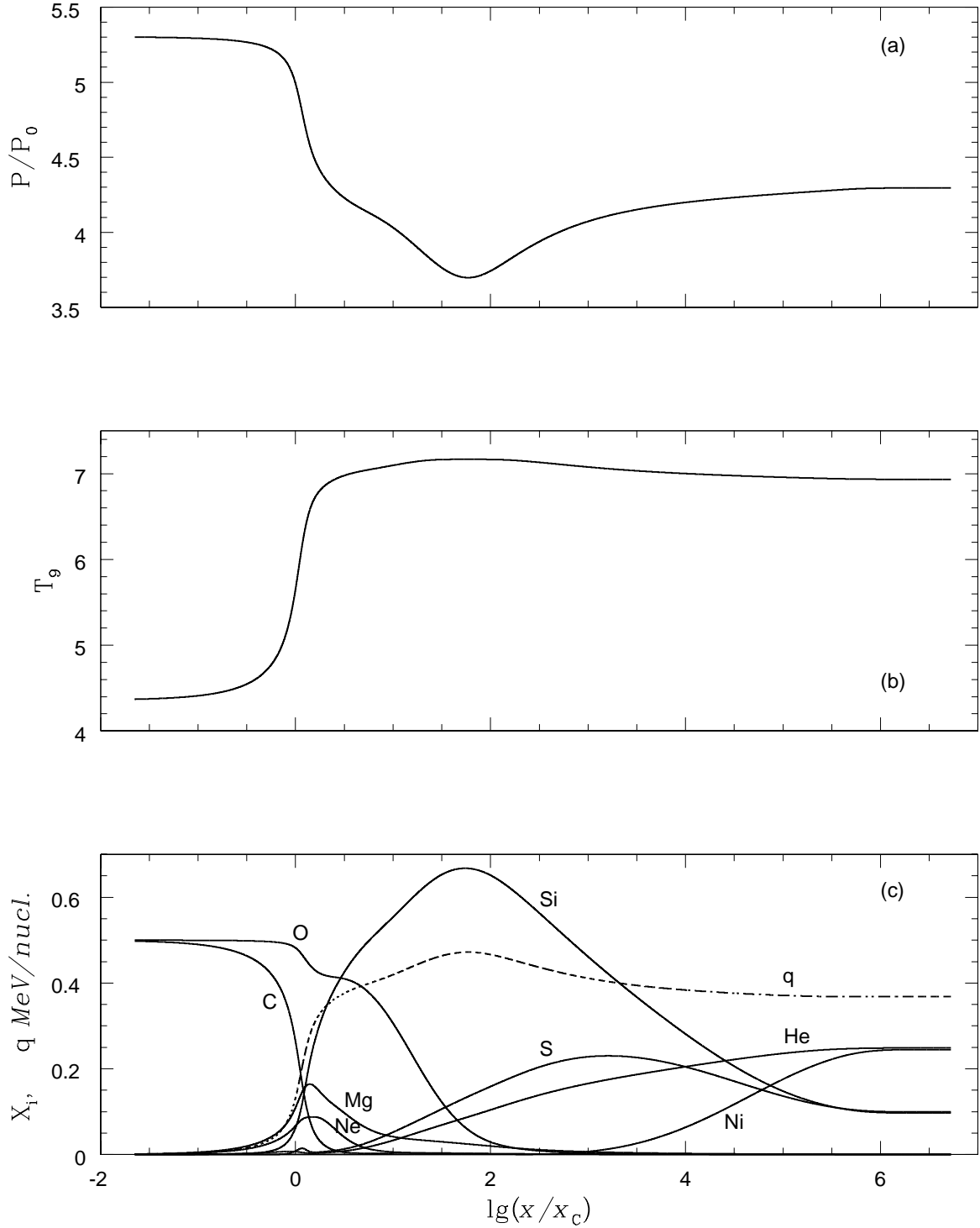

Fig. 3.— An overdriven detonation structure for $0.5C + 0.5O$ mixture at $\rho_0 = 10^8$ g/cm$^3$ and $f = 1.08$. Top - pressure, middle - temperature in $10^9$K, bottom - mass fractions of nuclei (solid curves) and nuclear energy release (dashed line). Distance is normalized to a half-reaction length of a C-burning layer. Shock wave is on the left (at $-\infty$ in logarithmic coordinates).

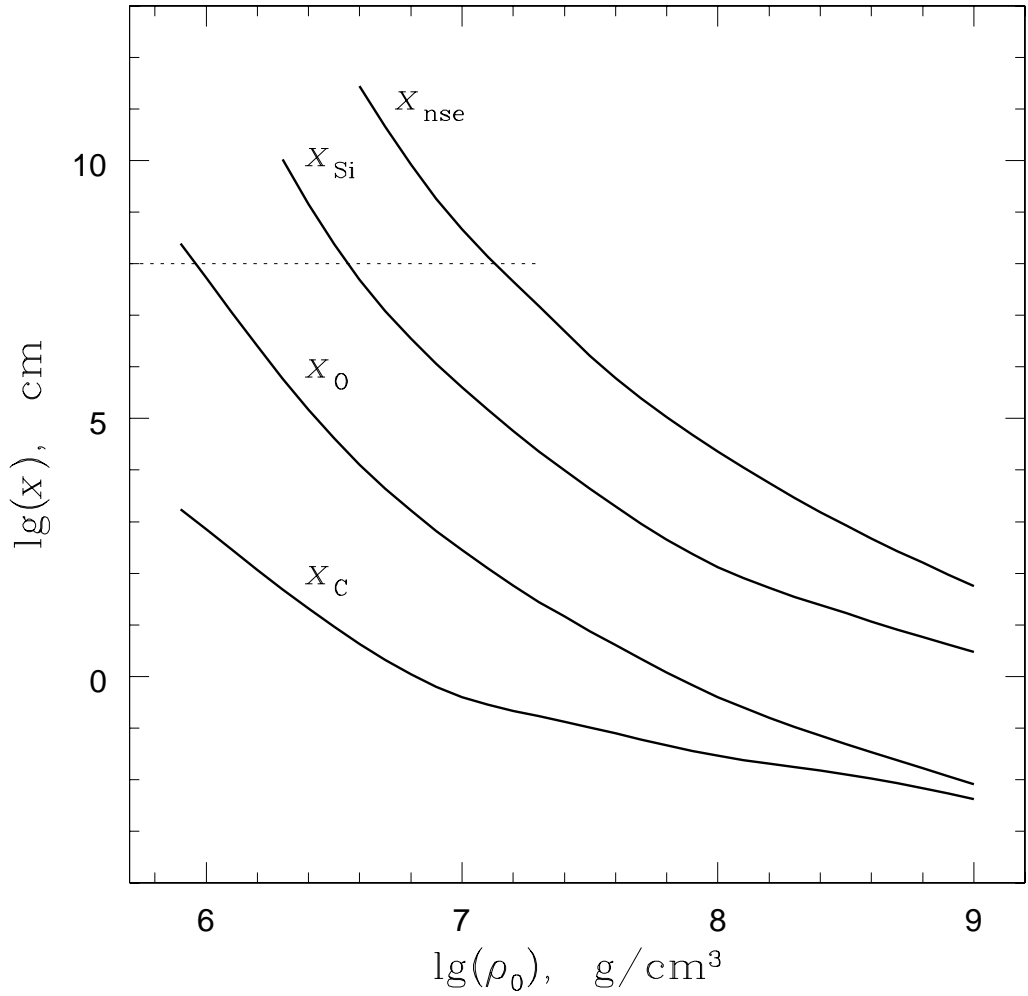

Fig. 4.— Thickness of reaction layers as a function of density for unsupported NSE detonations in $0.5C + 0.5O$ mixture. $x_C$ - half-reaction thickness of $C$-burning layer, $x_O$ - half-reaction thickness of $O$-burning layer, $x_{Si}$ - half-reaction thickness of $Si$-burning layer, $x_{NSE}$ - total thickness of the detonation wave.

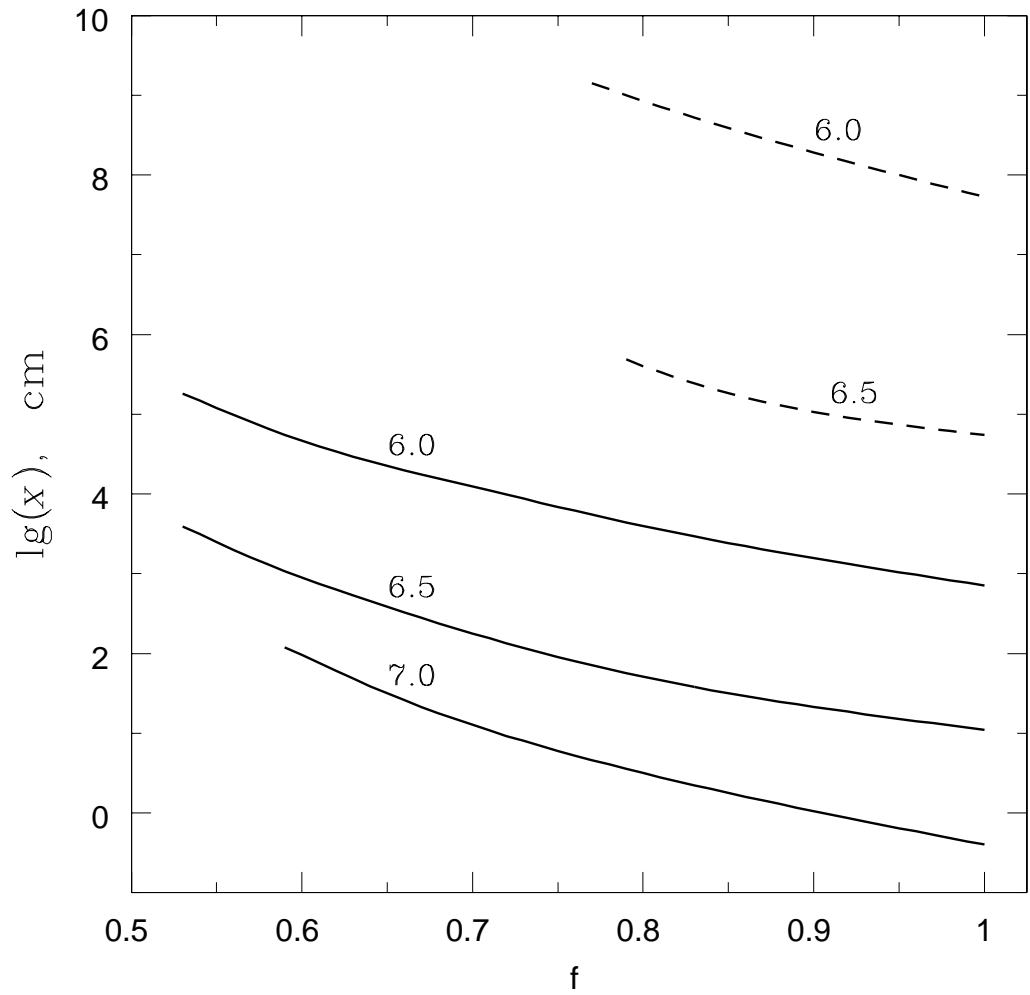

Fig. 5.— Thickness of C-detonation ($x_C$, solid lines) and O-detonation ($x_O$, dashed lines) as a function of overdrive $f \leq 1$ for densities $lg(\rho_0) = 6.0, 6.5, 7.0$; $\rho_0$ is in g/cm$^3$.

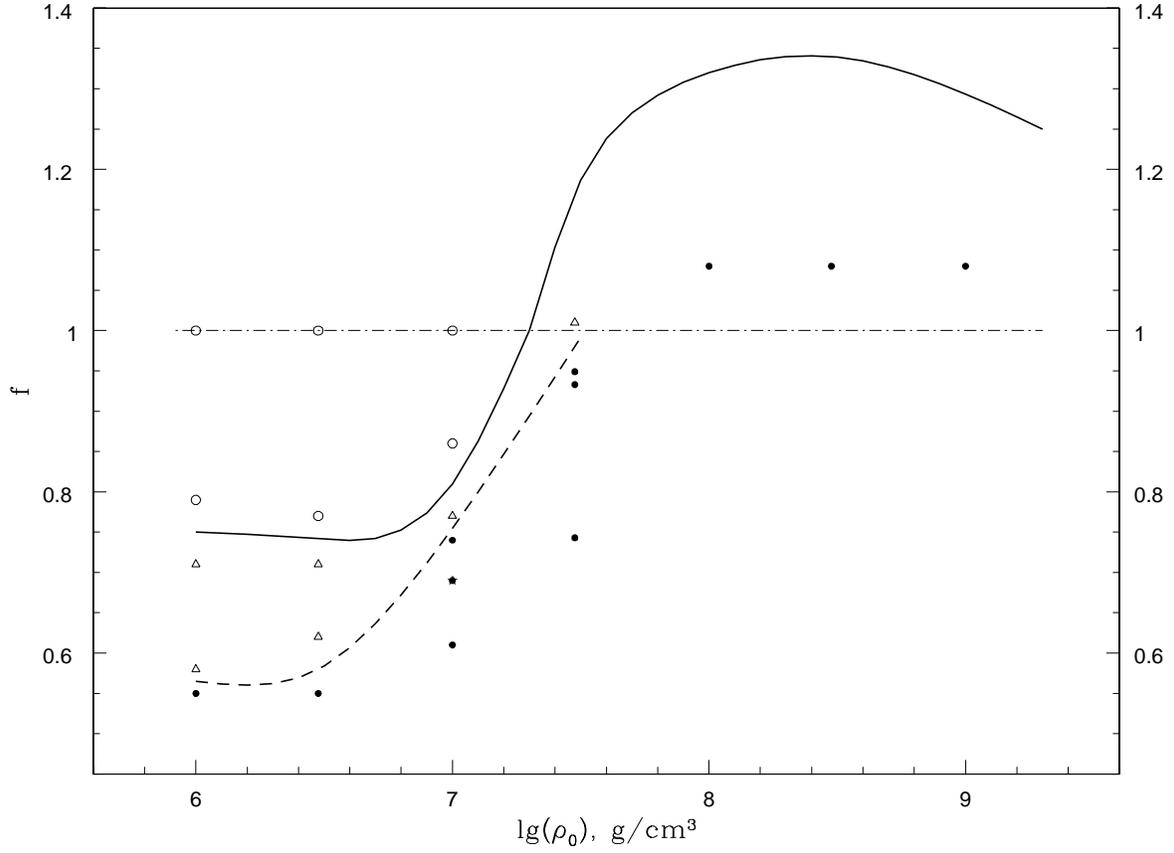

Fig. 6.— Summary of calculations of time-dependent propagation in $0.5C + 0.5O$ mixtures. Open circles - stable detonations; Open triangles - unstable oscillating detonations; solid triangles - unstable detonations for which the oscillation cycle was not calculated. Solid line - the estimated stability curve. Dashed line - time-scale of the oscillation cycle exceeds $\simeq 200 t_C$ where $t_C$ is the half-reaction timescale of a steady-state $C$-detonation. Dash-dotted line - location of Chapman-Jouguet detonations, $f = 1$.

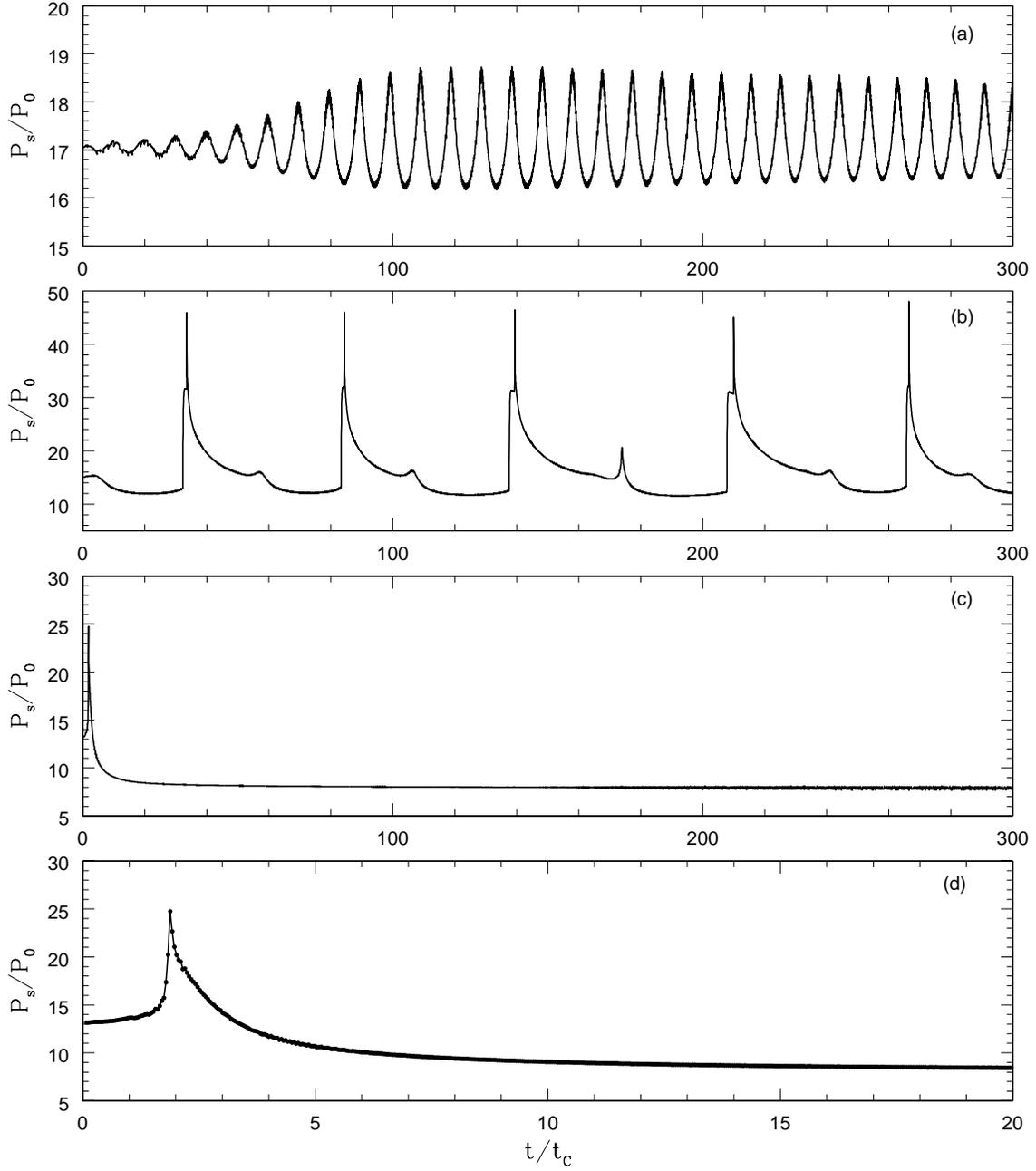

Fig. 7.— Post-shock pressure $P_s(t)$ for C-detonations with different overdrives at $\rho_0 = 3 \times 10^6$ g/cm$^3$. (a) $f = 0.71$; (b) $f = 0.62$, and (c) $f = 0.55$. (d) is the early portion of (c) and compares two simulations with different distance between the shock and the outflow boundary (dots - shorter distance; see text). Time is normalized to a half-reaction timescale of a steady-state detonation.

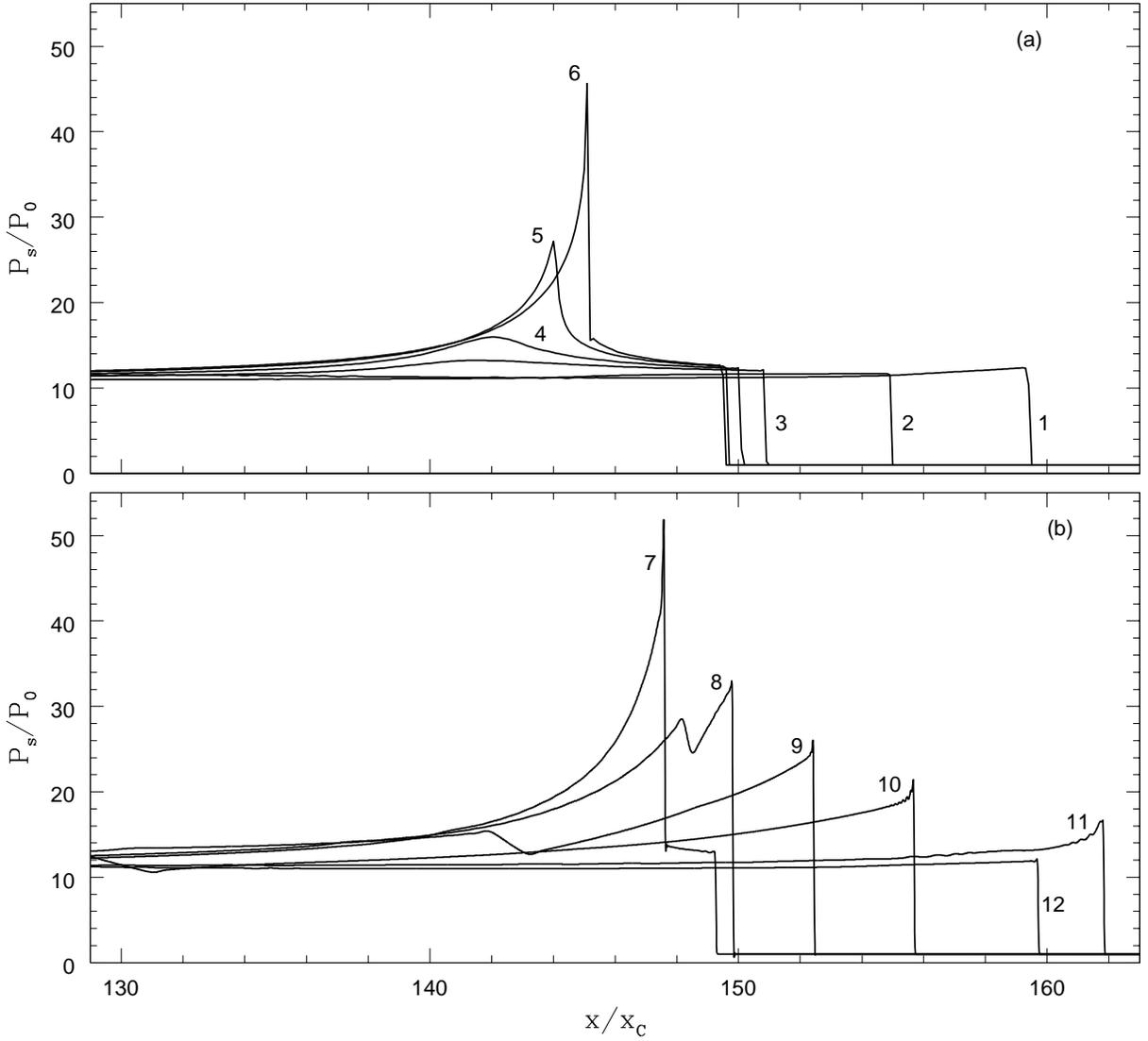

Fig. 8.— Oscillation cycle of the $f = 0.62$ C-detonation at $\rho_0 = 3 \times 10^6$ g/cm$^3$ shown in Fig. 7b. Plots of pressure, $P(x)$, for various moments of time during the third cycle. Distance is normalized to a half-reaction length-scale of a steady-state detonation. Times are (1) - 113.2, (2) - 124.3, (3) - 134.1, (4) - 136.2, (5) - 137.5, (6) - 137.9, (7) - 138.8, (8) - 139.2, (9) - 139.6, (10) - 142.2, (11) - 174.5, (12) - 183.1; normalized to a half-reaction time-scale of a steady-state detonation. (a) - formation of the spontaneous reaction wave and the secondary detonation in post-shock matter behind the weakening leading shock (times 1 to 6); (b) - re-establishment and subsequent weakening of the primary detonation (times 7 - 12).

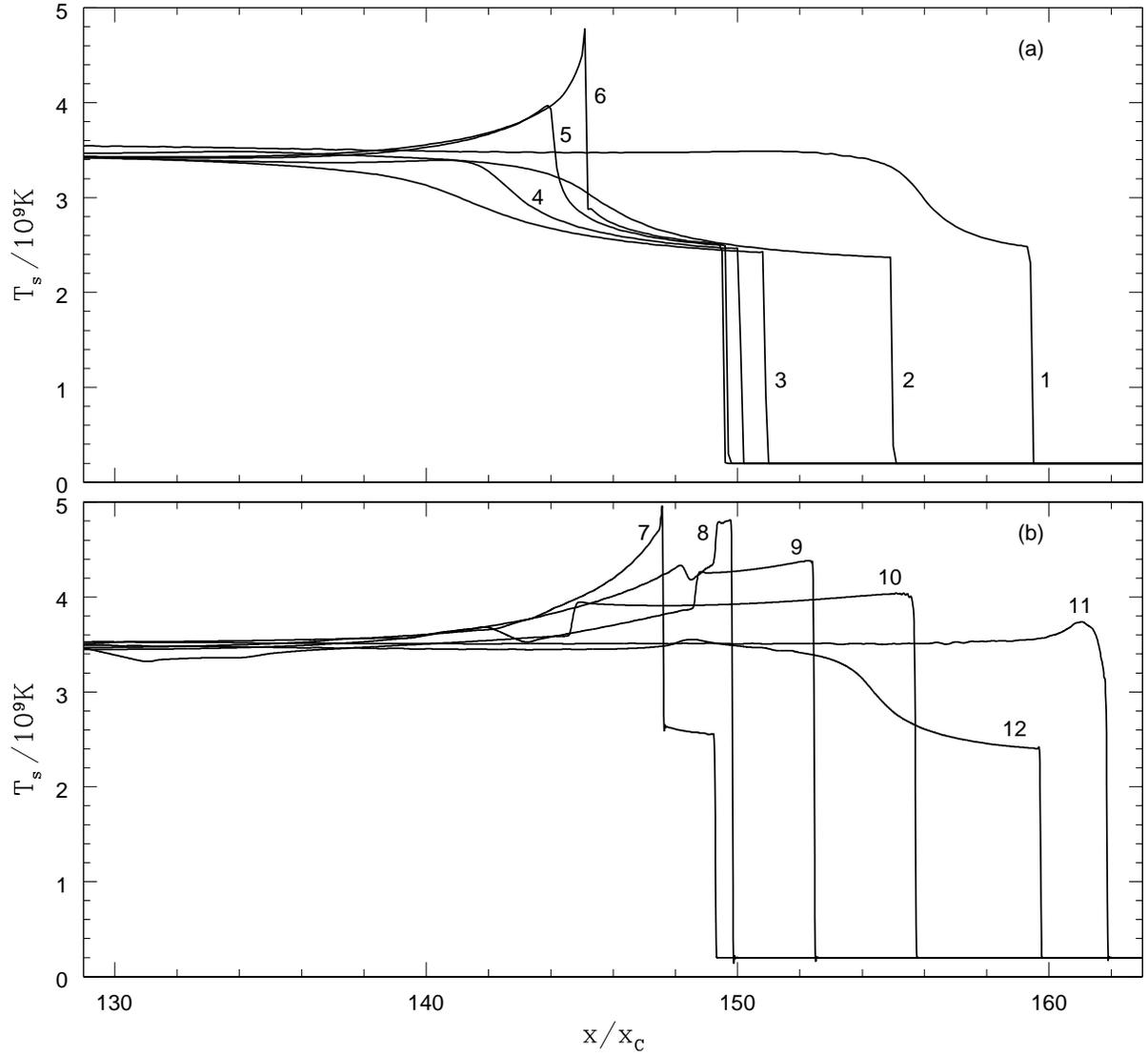

Fig. 9.— Oscillation cycle of the $f = 0.62$ C-detonation at $\rho_0 = 3 \times 10^6$ g/cm$^3$. Same as Fig. 8 but shows plots of temperature $T(x)$.

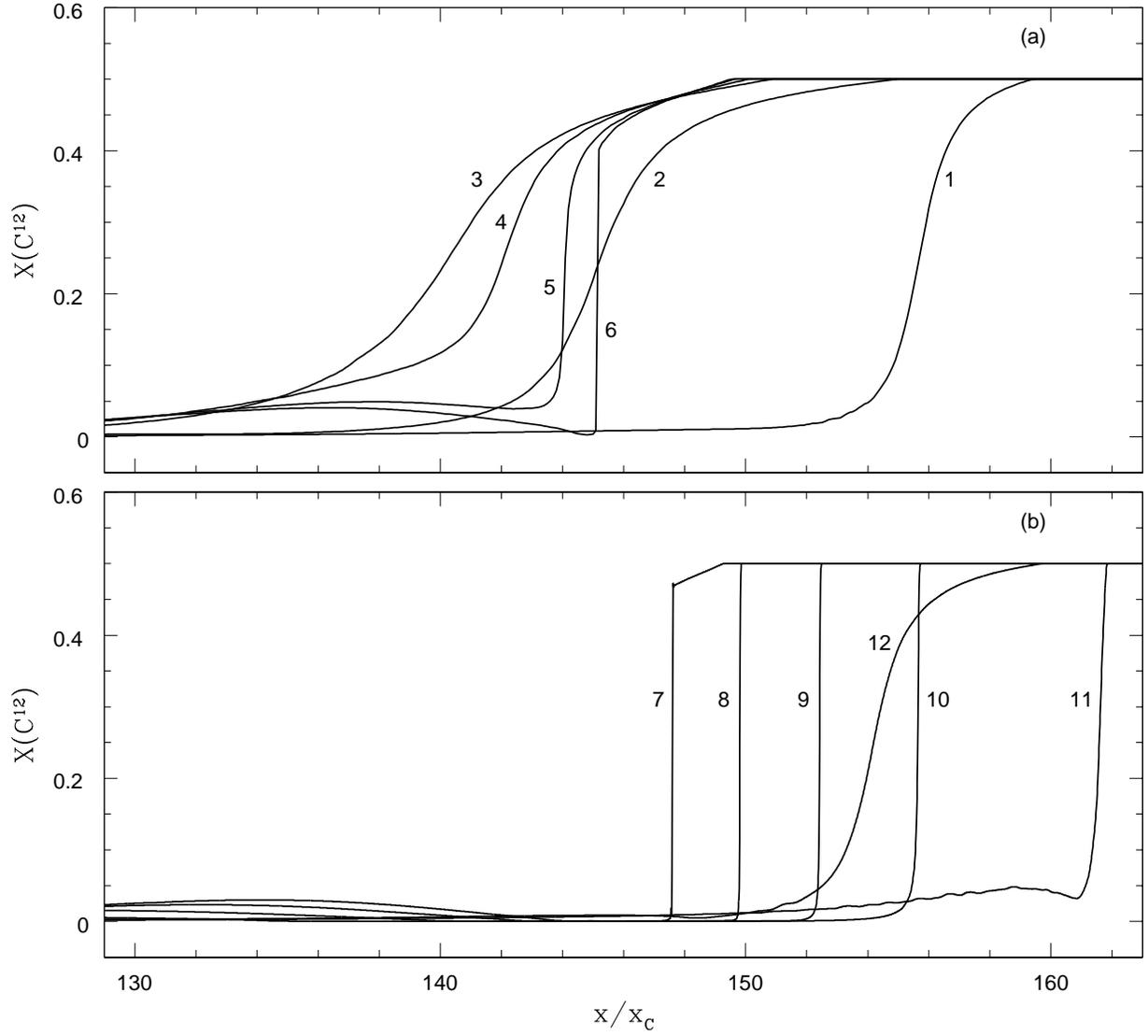

Fig. 10.— Oscillation cycle of the $f = 0.62$ C-detonation at $\rho_0 = 3 \times 10^6$ g/cm$^3$. Same as Fig. 8 but shows plots of carbon mass fraction $X(C^{12})$.

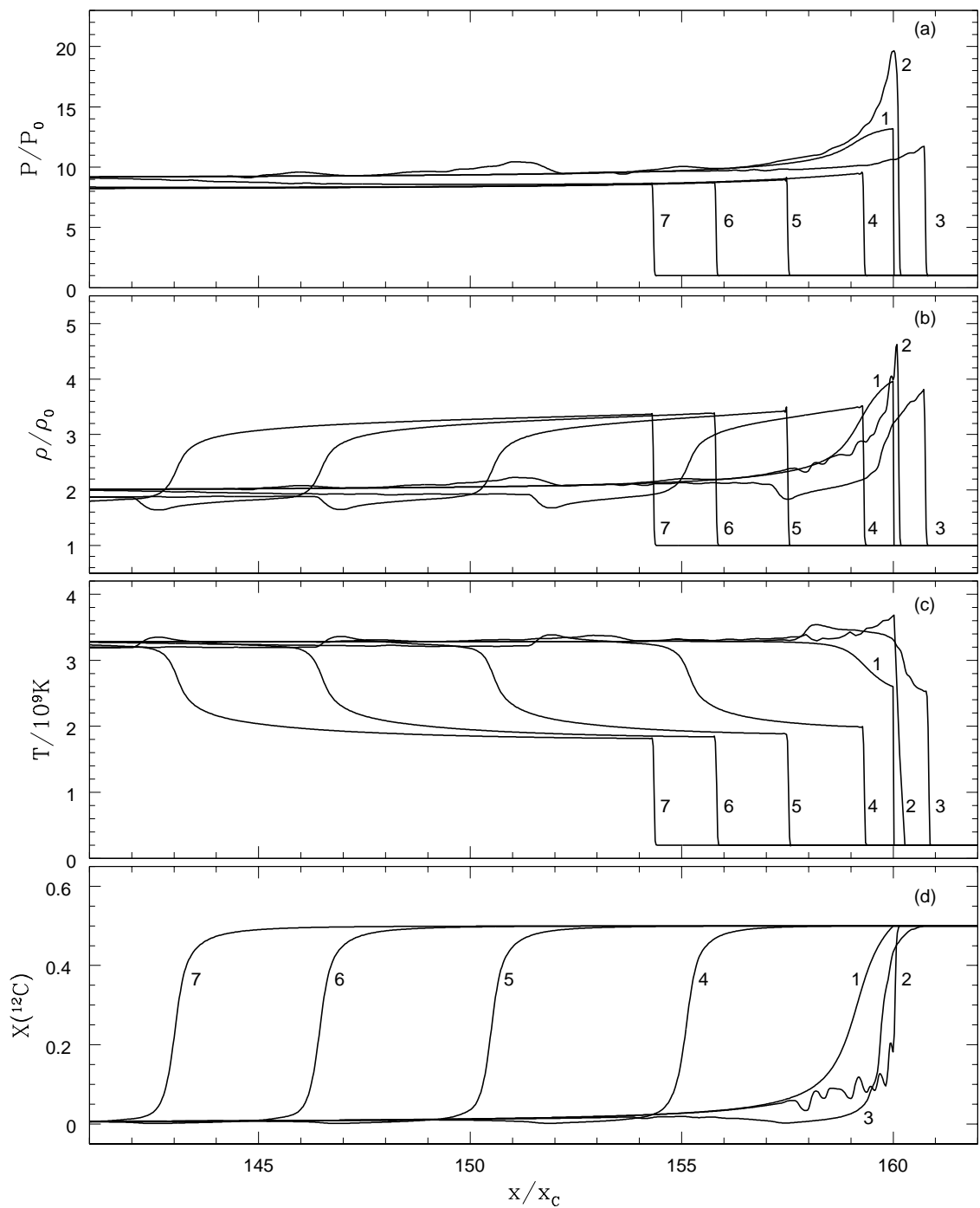

Fig. 11.— Propagation of a detonation wave for $\rho_0 = 3 \times 10^6$ g/cm$^3$ and $f = 0.55$. The figure shows the development of the instability and separation of the reaction zone from the leading shock. Distance is normalized to a half-reaction length-scale of a steady-state detonation. Times are (1) - 115.6, (2) - 134.1, (3) - 136.2, (4) - 137.5, (5) - 137.9, (6) - 138.4, (7) - 138.8; normalized to a half-reaction time-scale of a steady-state detonation.

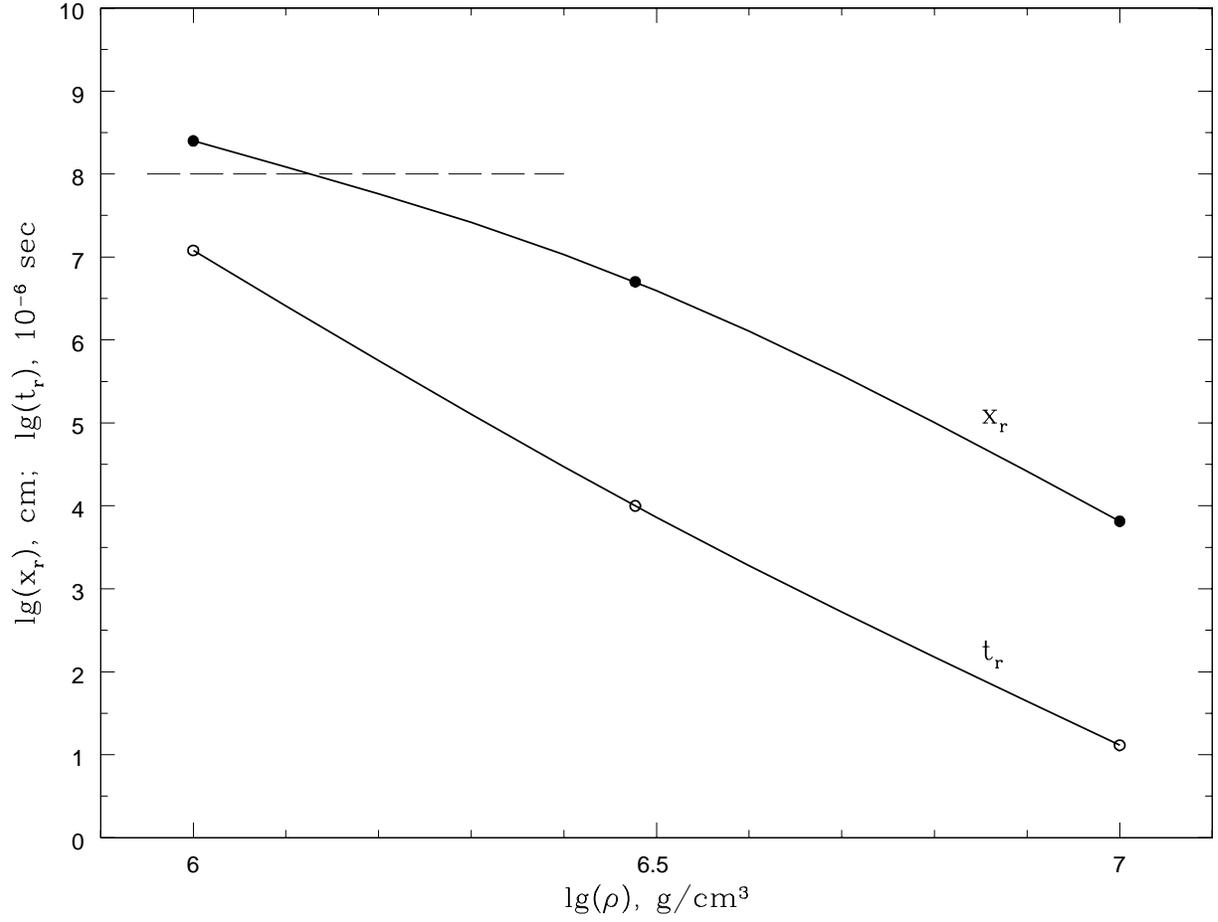

Fig. 12.— Solid circles - estimated reignition length-scale; Open circles - estimated reignition time-scale (see Sect. 4.3). Horizontal dashed line - characteristic spatial scale of an exploding Chandrasekhar-mass CO white dwarf.

– 36 –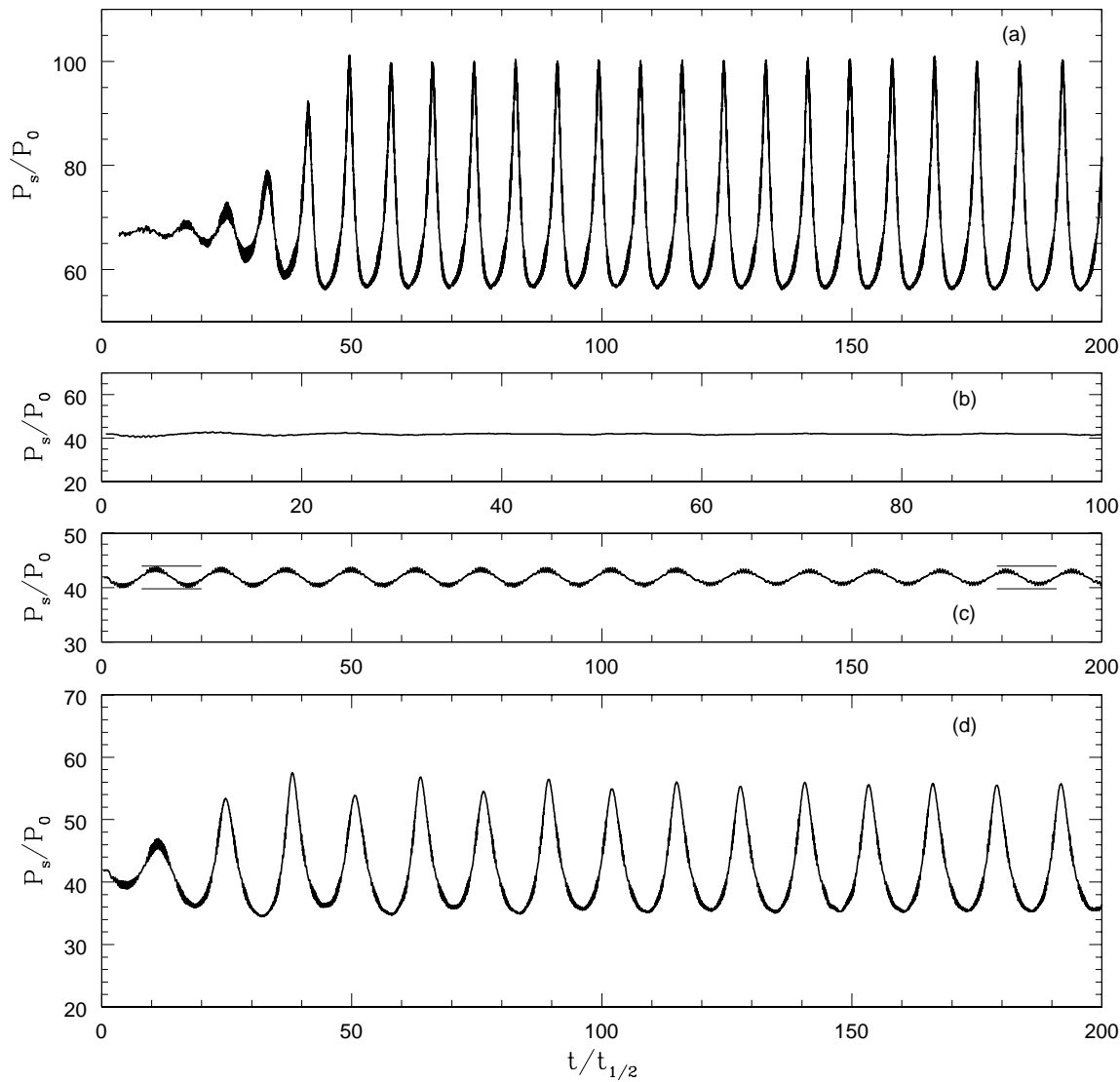

Fig. 13.— Post-shock pressure $P_s$ as a function of time for Arrhenius detonations. (a) - an overdriven detonation with $Q = q = 50$ and $f = 1.6$. The last three frames show results for a CJ detonation with $f = 1$, $q = 50$, and varying activation energy. (b) - $Q = 24$, (c) - $Q = 25$, and (d) - $Q = 27$. Time is normalized to a half-reaction time-scale of a steady-state detonation.